\documentclass[useAMS,usenatbib]{mn2e}
\pdfoutput=1

\usepackage{graphicx}
\usepackage{natbib}
\usepackage[figuresleft]{rotating}

\newcommand{\HII}{H{\,\small II}}

\newcommand{\ltsima} {$\; \buildrel < \over \sim \;$}
\newcommand{\gtsima} {$\; \buildrel > \over \sim \;$}
\newcommand{\lta} {\lower.5ex\hbox{\ltsima}}
\newcommand{\gta} {\lower.5ex\hbox{\gtsima}}

\newcommand{\kms}{km\ s$^{-1}$}

\newcommand{\lya}{Ly$\alpha$}
\title[CO(1-0) survey of high-z radio galaxies]{CO(1-0) survey of high-$z$ radio galaxies: alignment of molecular halo gas with distant radio sources\thanks{Based on Australia Telescope Compact Array observations}}
\author[B.\,H.\,C. Emonts et al.]{B. H. C. Emonts$^{1,2}$\thanks{E-mail:bjornemonts@gmail.com}, R. P. Norris$^{2}$, I. Feain$^{3}$, M. Y. Mao$^{4}$, R. D. Ekers$^{2}$, G. Miley$^{5}$,        
\newauthor N. Seymour$^{2}$, H. J. A. R\"{o}ttgering$^{5}$, M. Villar-Mart\'{i}n$^{1}$, E. M. Sadler$^{3}$, C. L. Carilli$^{4}$,   
\newauthor E. K. Mahony$^{6}$, C. de Breuck$^{7}$, A. Stroe$^{5}$, L. Pentericci$^{8}$, G. A. van Moorsel$^{4}$,
\newauthor G. Drouart$^{7}$, R. J. Ivison$^{9}$, T. R. Greve$^{10}$, A. Humphrey$^{11}$, D. Wylezalek$^{7}$,
\newauthor C. N. Tadhunter$^{12}$\\
$^{1}$Centro de Astrobiolog\'{i}a (INTA-CSIC), Ctra de Torrej\'{o}n a Ajalvir, km 4, 28850 Torrej\'{o}n de Ardoz, Madrid, Spain\\
$^{2}$CSIRO Astronomy and Space Science, Australia Telescope National Facility, PO Box 76, Epping NSW, 1710, Australia\\
$^{3}$School of Physics, University of Sydney, NSW 2006, Australia\\
$^{4}$National Radio Astronomy Observatory, P.O. Box 0, Socorro, NM 87801-0387, USA\\
$^{5}$Leiden Observatory, University of Leiden, P.O. Box 9513, 2300 RA Leiden, Netherlands\\
$^{6}$Netherlands Institute for Radio Astronomy, Postbus 2, 7990 AA Dwingeloo, the Netherlands\\
$^{7}$European Southern Observatory, Karl Schwarzschild Strasse 2, 85748 Garching, Germany\\
$^{8}$INAF Osservatorio Astronomico di Roma, Via Frascati 33,00040 Monteporzio (RM), Italy\\
$^{9}$Institute for Astronomy, University of Edinburgh, Blackford Hill, Edinburgh EH9 3HJ\\ 
$^{10}$Department of Physics and Astronomy, University College London, Gower Street, London WC1E 6BT, UK\\
$^{11}$Centro de Astrof\'{i}sica da Universidade do Porto, Rua das Estrelas, 4150-762 Porto, Portugal\\
$^{12}$Department of Physics and Astronomy, University of Sheffield, Sheffield S3 7RH, UK
}
\begin{document}

\date{}

\pagerange{\pageref{firstpage}--\pageref{lastpage}} \pubyear{2013}

\maketitle

\label{firstpage}

\begin{abstract}
We present a CO(1-0) survey for cold molecular gas in a representative sample of 13 high-$z$ radio galaxies (HzRGs) at 1.4\,$<$\,$z$\,$<$\,2.8, using the Australia Telescope Compact Array. We detect CO(1-0) emission associated with five sources: MRC\,0114-211, MRC\,0152-209, MRC\,0156-252, MRC\,1138-262 and MRC\,2048-272. The CO(1-0) luminosities are in the range $L'_{\rm CO} \sim (5 - 9) \times 10^{10}$\,K\,\kms\,pc$^{2}$. For MRC\,0152-209 and MRC\,1138-262 part of the CO(1-0) emission coincides with the radio galaxy, while part is spread on scales of tens of kpc and likely associated with galaxy mergers. The molecular gas mass derived for these two systems is ${\rm M}_{\rm H2} \sim 6 \times 10^{10}\, {\rm M}_{\odot}$ (M$_{\rm H2}$/$L'_{\rm CO}$\,=\,0.8). For the remaining three CO-detected sources, the CO(1-0) emission is located in the halo ($\sim$50-kpc) environment. These three HzRGs are among the fainter far-IR emitters in our sample, suggesting that similar reservoirs of cold molecular halo gas may have been missed in earlier studies due to pre-selection of IR-bright sources. In all three cases the CO(1-0) is aligned along the radio axis and found beyond the brightest radio hot-spot, in a region devoid of 4.5$\mu$m emission in Spitzer imaging. The CO(1-0) profiles are broad, with velocity widths of $\sim$ 1000 - 3600 \kms. We discuss several possible scenarios to explain these halo reservoirs of CO(1-0). Following these results, we complement our CO(1-0) study with detections of extended CO from the literature and find at marginal statistical significance (95$\%$ level) that CO in HzRGs is preferentially aligned towards the radio jet axis. For the eight sources in which we do not detect CO(1-0), we set realistic upper limits of $L'_{\rm CO} \sim 3-4 \times 10^{10}$ K\,\kms\,pc$^{2}$. Our survey reveals a CO(1-0) detection rate of 38$\%$, allowing us to compare the CO(1-0) content of HzRGs with that of other types of high-$z$ galaxies.
\end{abstract}


\section{Introduction}
\label{sec:intro}

High-redshift radio galaxies (HzRGs, with $L_{500\,MHz} > 10^{27}$ W\,Hz$^{-1}$) are among the most massive and best studied galaxies in the early Universe \citep{sey07,bre10}. They have traditionally been identified by the ultra-steep spectrum of their easily detectable radio continuum, which served as a beacon for tracing the faint host galaxy environment \citep{rot94,cha96}. HzRGs have been observed to be the signposts of large overdensities in the early Universe, the so-called proto-clusters that are believed to be the ancestors of local rich clusters \citep[e.g.][]{ven07,mil08,wyl13}. The radio host galaxies are typically the massive central sources of these proto-clusters and are surrounded by giant (100\,kpc-scale) ionised gas halos \citep{vil02,vil03,vil06,vil07,hum07,hum13}. A significant fraction of these gaseous emission-line halos show spatially resolved absorption from extended regions of neutral gas \citep{oji97lya,jar03,hum08}. 

HzRGs are in a very active stage of their evolution. They show clumpy optical morphologies \citep{pen01}, which indicates that continuous mergers are taking place. They often also contain evidence for massive star formation \citep[e.g.][]{dun94,ivi95,ivi00,arc01,ste03,sey12,roc13}. Alignments have been seen between the radio synchrotron jets that emanate from the central super-massive black-hole and optical/UV emission from warm gas and stellar continuum \cite[e.g.][]{mcc87,cha87}, as well as X-ray emission \citep[e.g.][]{car02,sma13} and dust re-radiated submillimetre emission \citep{ste03}. The powerful radio jets can also exert significant feedback onto the surrounding inter-stellar medium \citep[ISM; e.g.][]{vil03,nes06,hum06,ogl12}. On average, HzRGs with smaller jets show stronger jet-ISM interaction, more intense star formation and larger reservoirs of neutral gas \citep{oji97,hum06,hum11}.

A crucial component in the research on high-z proto-cluster radio galaxies is the study of cold molecular gas, which is the raw ingredient for star formation (and potential fuel for the AGN). An excellent review on the research of cold gas in high-$z$ galaxies is given by \citet{car13}. Because molecular hydrogen (H$_{2}$) has strongly forbidden transitions, it can only be detected directly when it is shocked-heated to high (T$>$100\,K) temperatures. An excellent tracer for the cold component of molecular gas is carbon monoxide (CO), because it is the most abundant molecule after H$_{2}$ and its excitation to the various $^{12}$CO({\sl J,}{\sl J\,-\,1}) transitions occurs at low temperatures, through collisions with H$_{2}$ \citep[even at modest densities;][]{sol05}. 

In the 1990's, the first surveys for CO in HzRGs (mainly using the higher $J$-transitions) did not find any cold molecular gas \citep{eva96,oji97}. Subsequently, improvements in millimetre receivers brought interesting results on CO in individual HzRGs between $z \sim 2 - 5$. \citep[][see also review by \citealt{mil08}]{sco97,all00,pap00,pap01,pap05,bre03,bre03AR,bre05,gre04,kla05,ivi08,ivi12,nes09,emo11a,emo13}. In some cases CO is resolved on tens of kpc scales \citep{ivi12}, associated with various components \citep[e.g. merging gas-rich galaxies;][]{bre05,emo13}, or found in giant Ly$\alpha$ halos that surrounds the host galaxy \citep{nes09}. Several studies also identified alignments between the CO emission and the radio jet axis \citep{kla04,nes09}. These results show that detectable amounts of cold molecular gas in HzRGs are not restricted to the central region of the radio galaxy. 

Despite these interesting results, a lack of sensitivity and bandwidth coverage at existing millimetre facilities (often not more than the width of the CO line, or the accuracy of the redshift) severely hindered systematic searches for molecular gas in HzRGs. Instead, targets were generally pre-selected on a high infra-red (IR) of submillimetre (submm) luminosity.

The introduction of broad-bandwidth receivers at most of the large millimetre observatories has opened new possibilities for accurate searches for CO at high-z. A crucial species for quantifying the cold molecular gas is the ground transition CO(1-0). While the high CO transitions trace dense and thermally excited gas in the central starburst/AGN region, only the lowest CO transitions may fully reveal widely distributed reservoirs of less dense, sub-thermally excited gas \citep[e.g.][]{pap00,pap01,car10,ivi11}. The ground transition CO(1-0) is least affected by the excitation conditions of the gas, hence observations of CO(1-0) provide the most robust estimates of the overall molecular gas content.\footnote{Although see \citet{pap12} for important caveats when using only CO(1-0) to determine the molecular gas content.} For high-$z$ systems, observations of CO(1-0) require observing capabilities in the 20-50\,GHz regime, hence accurate studies of CO(1-0) at $z>1.3$ have become feasible with the vastly improved millimetre receivers of the Karl G. Jansky Very Large Array (JVLA) and Australia Telescope Compact Array (ATCA) \citep[e.g.][]{ara10,emo11b}. Only a handful of observations of CO(1-0) in HzRGs currently exist, but they show evidence for large gas reservoirs \citep[in some cases spread across tens of kpc;][]{ivi12,emo11a,emo13}.

In this paper, we present a survey for CO(1-0) emission in a representative sample of 13 high-z radio galaxies with the ATCA. Throughout this paper we will assume $H_{0} = 71$\,\kms\,Mpc$^{-1}$, $\Omega_{\rm M} = 0.27$ and $\Omega_{\Lambda} = 0.73$.

\subsection{Sample}

We initially selected those HzRGs from the flux-limited 408\,MHz Molonglo Reference Catalogue \citep{lar81} which were defined by \citet{mcc96} to have $S_{\rm 408\,MHz} > 0.95$\,Jy, $-30^{\circ} < {\rm dec} < -20^{\circ}$ and $9^{\rm h}20^{\rm m} < {\rm R.A.} < 14^{\rm h}4^{\rm m}$ or $20^{\rm h}20^{\rm m} < {\rm R.A.} < 6^{\rm h}14^{\rm m}$\footnote{This approach primarily selects sources based on their flux density, not their ultra-steep spectrum, as in other effective searches for HzRGs \citep[][]{rot94,cha96}}. We subsequently selected for observing those sources for which the CO(1-0) line ($\nu_{\rm rest} = 115.271203$\,GHz) is observable in the ATCA 7mm band ($30-50$\,GHz), corresponding to a redshift range of 1.3$<z_{\rm CO(1-0)}<$2.8. In addition, in order to keep the sample manageable (given the large amount of observing time required; Sect.\,\ref{sec:observations}) and to maximize the scientific output, we only included sources for which complementary optical and infra-red data from the {\sl Hubble Space Telescope (HST)} and {\sl Spitzer Space Telescope} are available \citep{sey07,bre10,pen01}. The well-studied source MRC\,0211-112 \citep[outside the declination range sampled by][]{mcc96} was also added to our sample \citep{oji97lya,pen01,ver01,hum13,bre10}

In total, our sample consists of 13 southern MRC sources (see Table \ref{tab:obs}). All 13 sources have a 3\,GHz rest-frame radio luminosity of $L_{\rm 3\,GHz} > 10^{27.5}$\,W\,Hz$^{-1}$ and they sample the full range in mid-IR (5 $\mu$m restframe) luminosities found across a large sample of HzRGs by \citet{bre10}. Our sources also sample the K-band magnitude range covered by \citet{bre10} for objects with $16 < {\rm mag}_{K} \le 19$, but our sample does not include sources in their lowest mag$_{K}$ bin ($19 < {\rm mag}_{K} \le 20$). We argue that this is likely because of optical/near-IR selection biases in the existing {\sl HST/WFPC2} and {\sl HST/NICMOS} studies on which we based our sample selection.

An additional source (MRC\,0943-242; $z=2.92$) was observed in CO(1-0) just beyond the nominal edge of the ATCA 7mm band ($\nu_{\rm CO} = 29.4$\,GHz), where increased noise and low-level instrumental effects prevented us from reaching the same sensitivity as for the 13 sample sources presented in this paper. MRC\,0943-242 is therefore excluded from this paper. Results of MRC\,0943-242, including a tentative CO(1-0) detection in the halo of this HzRG, have been described in \citet{emo11b}.

\section{Observations}
\label{sec:observations}

CO(1-0) observations were performed with the Australia Telescope Compact Array (ATCA) during 2009 - 2013 in the most compact hybrid H75, H168 and H214 array configurations (which contain both an EW and NS spur and maximum baselines of 89, 192 and 247\,m respectively). The bulk of the observations were done during the night in the southern late-winter and early-spring (months of Aug, Sept, Oct). A summary of the observations is given in Table \ref{tab:obs}. 

The Compact Array Broadband Backend (CABB) with $2 \times 2$\,GHz receiver bands was used in its coarsest spectral resolution of 1\,MHz per channel \citep[][]{wil11}. For an object at $z=2$ at which the CO(1-0) emission line ($\nu_{\rm rest} = 115.271$\,GHz) is observed at $\nu_{\rm obs} = 38.400$\,GHz, this setup results in a velocity coverage of 16,000 \kms\ per 2\,GHz band with 8 \kms\ maximum spectral resolution. During the early observing period of this project (immediately after the CABB upgrade in 2009), several corrupted correlator blocks caused a significant number of channels to drop out in the first part of each 2\,GHz band. We therefore centred the bands such that the expected CO(1-0) signal would fall in the clean part of each band. During the later period this problem did not occur and we centred each band around the redshift of the expected CO(1-0) line. For redundancy (in case of technical issues, which occurred more frequently in the months after the CABB upgrade), both 2\,GHz bands were centred around the same frequency, but because the signal is not independent in the two bands, only a single band was used in the final analysis of each observation. To better handle potential bandpass or systematic effects, we frequently introduced a small offset (tens to hundreds of MHz) in the central frequency of the bands between runs on a particular target.

For data calibration, we followed the strategy described in \citet{emo11b}. The phases (and in many cases also the bandpass) were calibrated every 5-15 minutes with a short ($\sim$2 min) scan on a nearby bright calibrator (Table\,\ref{tab:obs}). In case the flux of this calibrator was $<$1 Jy, or the calibrator was not suitable for bandpass calibration, a strong bandpass calibrator was observed at the start, middle and end of each run. Fluxes were calibrated using (in order of preference) Uranus, Mars, PKS\,1934-638 or the ultra-compact \HII\ region G309 \citep[for the latter see][]{emo11b}. While the accuracy of the relative flux calibration between runs of the same target was $\la$\,5\,$\%$, the error in absolute flux calibration was $\sim$$20\%$, due to a small difference in the models for the various flux calibrators. We note, however, that the latter only introduces a scaling factor that depends on the used flux calibrator for a particular source.

For the data reduction and visualisation we used the software packages MIRIAD \citep{sau95} and KARMA \citep{goo96}. All of the HzRGs in our sample were detected in the 115\,GHz rest-frame radio continuum at the mJy level. The continuum was separated from the line data by fitting a straight line to the line-free channels in the uv-domain. Because of the large velocity coverage per 2\,GHz receiver band, the continuum subtraction could initially be done reliably using all channels, without affecting a potential weak CO signal. For those cases where CO(1-0) was detected, the continuum subtraction was repeated by excluding the channels in which the faint CO(1-0) signal was present. We note, however, that this did not significantly alter the final results. For MRC\,0114-211 the radio continuum flux is $\sim$80 mJy. This allowed us to do a self calibration on the continuum of MRC\,0114-211, which solutions were copied to the corresponding line data. The velocity axis of each individual line data set was transformed into optical barycentric velocity definition with respect to the redshift of the object (as given in Table \ref{tab:obs}). While each data set was reduced and analysed individually for CO signals, as well as for spurious signals or abnormalities, a Fourier transform was eventually performed on the combined line data for each object in our sample, using robust +1 weighting \citep{bri95}. Because of the low peak-flux of our CO(1-0) detections, no cleaning was done on the line data, except for a mild clean of the strong CO(1-0) signal in MRC\,0152-209 (Sect.\,\ref{sec:results}). The line data were subsequently binned to the results given in Table \ref{tab:results2} and Figs. \ref{fig:COprofiles} and \ref{fig:COupper}. Total intensity images were made by summing the signal across the channels in which CO(1-0) was detected (without setting a noise threshold/cutoff). Values of $L'_{\rm CO(1-0)}$ from the CO(1-0) detections provided in this paper have been derived from these total intensity images. The beam-size and rms noise level of our data is summarised in Table \ref{tab:results2}. The primary beam of our observations is 80 arcsec at $\nu_{\rm obs} = 35$\,GHz. 

\begin{figure*}
\centering
\includegraphics[width=0.92\textwidth]{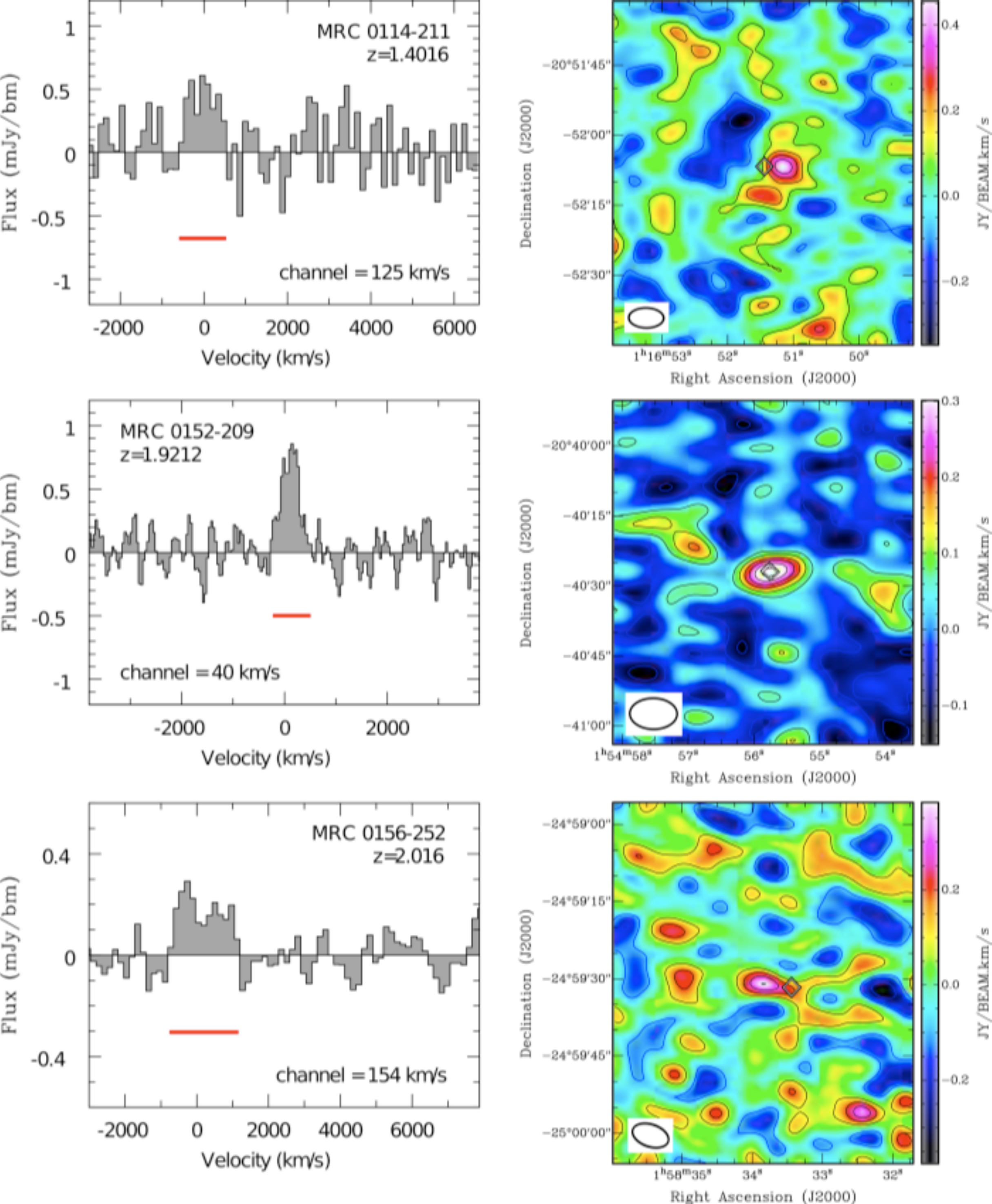}
\caption{CO(1-0) spectra and total intensity images of the five CO-detected sample sources: MRC\,0114-211, MRC\,0152-209 \citep{emo11a}, MRC\,0156-252, MRC\,1138-262 \citep[][]{emo13} and MRC\,2048-272. The channel width of the plotted spectra is shown at the bottom of each plot and was chosen to best visualize the CO(1-0) detection. For MRC\,0152-209, MRC\,0156-252 and MRC\,1138-262 a Hanning smoothed version of the spectra is shown to best visualise the CO(1-0) emission, while for MRC\,0114-211 and MRC\,2048-272 no Hanning smoothing was applied to better distinguish the CO(1-0) emission from the noise characteristics across the bandpass. The 0-velocity is defined as the optical redshift of the host galaxy \citep[see][and references therein]{bre00}, except for MRC\,0114-211, where the uncertainty in optical redshift is as large as the velocity coverage of our CO observations (see Appendix \ref{sec:app}). The red bar indicates the velocity range across which we integrated to obtain the total intensity map of each source. Contours levels are spaced 1$\sigma$ apart (with $\sigma = 0.094, 0.095, 0.095, 0.046, 0.080$ Jy\,beam$^{-1}$\,$\times$\,\kms\ for MRC\,0114-211, MRC\,0152-209, MRC\,0156-252, MRC\,1138-262, MRC\,2048-272, respectively). The diamond indicates the position of the host galaxy. {\sl [Color versions of all the Figures in this paper are available in the on-line edition.]}}
\label{fig:COprofiles}
\end{figure*}

\addtocounter{figure}{-1}
\begin{figure*}
\centering
\includegraphics[width=0.92\textwidth]{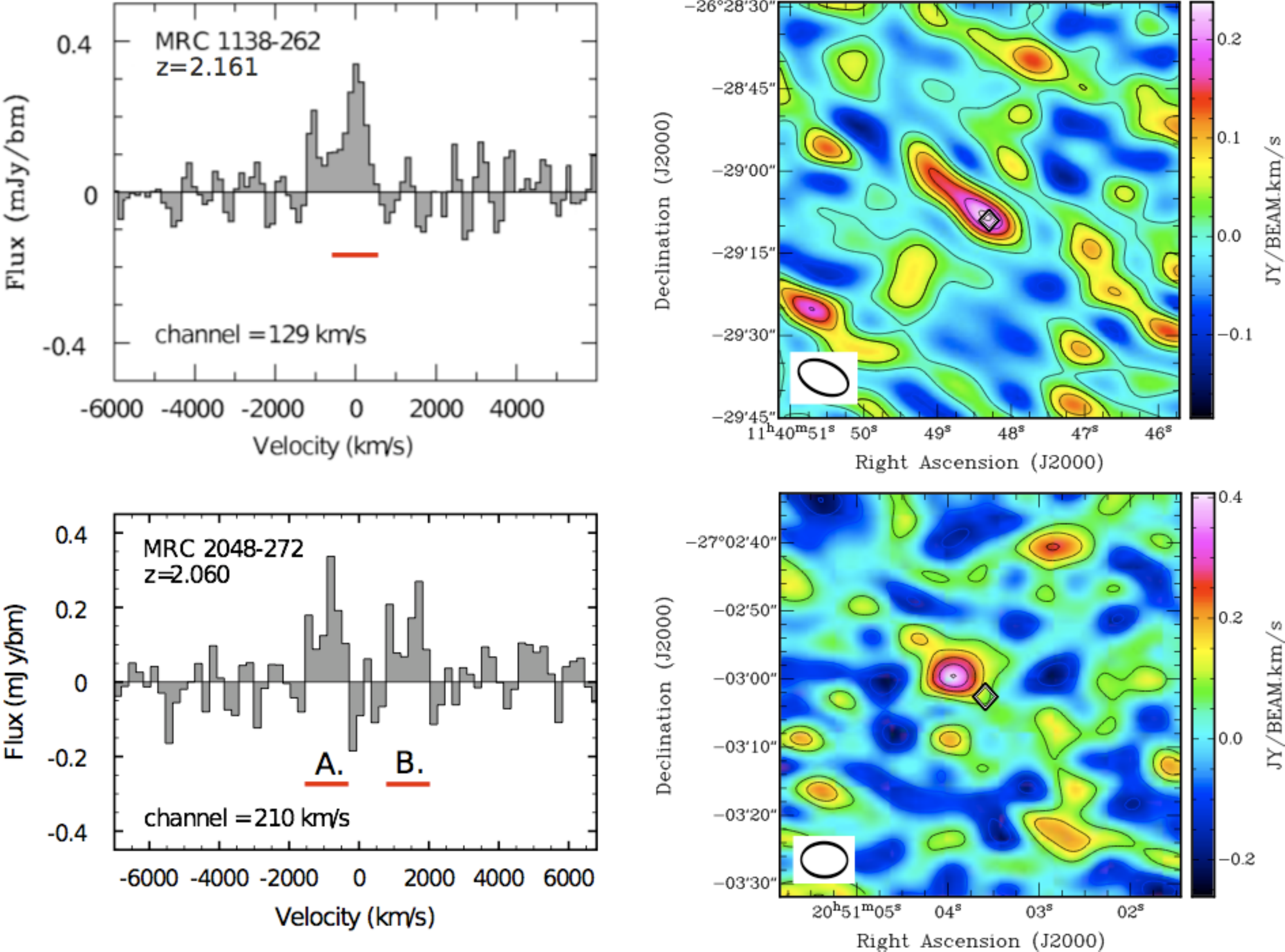}
\caption{-- {\sl continued}}
\end{figure*}

The continuum data for each object were combined into a map of the 115\,GHz restframe emission. By comparing the location of the peak emission (and in some cases also the morphological structure) of this radio continuum with published high-resolution 4.7 and 8.2\,GHz data \citep{car97,pen00radio,bre10}, we were able to verify that the positional accuracy of our mm data (line and continuum) is better than 1 arcsec. The presentation of the continuum data is left for a future paper.

\section{Results}
\label{sec:results}

\subsection{CO detections and upper limits}
\label{sec:profiles}

We detect CO(1-0) emission associated with five of our sample sources. These are MRC\,0114-211 ($z=1.40$), MRC\,0152-209 ($z=1.92$), MRC\,0156-252 ($z=2.02$), MRC\,1138-262 (also called the `Spiderweb Galaxy'; $z=2.16$) and MRC\,2048-272 ($z=2.06$). The CO(1-0) emission-line profiles of these five sources are shown in Fig.\,\ref{fig:COprofiles} and their properties are summarised in Table\,\ref{tab:results2}. The characteristics and spatial distribution of the CO(1-0) emission in these sources will be discussed in Sect.\,\ref{sec:characteristics}.

The CO(1-0) emission-line luminosities of the five CO(1-0) detected sources in Fig.\,\ref{fig:COprofiles} are in the range $L'_{\rm CO} = 4.5 - 9.2 \times 10^{10}$\,${\rm K~km~s^{-1}~pc^2}$ (see Table\,\ref{tab:results2}). $L'_{\rm CO}$ was calculated following \citet[][and references therein]{sol05}:
\ \\
\begin{equation}
L'_{\rm CO} = 3.25 \times 10^7\left(\frac{\int_{\rm v} S_{\rm CO} \delta {\rm v}}{{\rm Jy}~{\rm km/s}}\right)\left(\frac{D_{\rm L}}{{\rm Mpc}}\right)^2\left(\frac{{\nu_{\rm rest}}}{\rm GHz}\right)^{-2}\left(1+z\right)^{-1},
\label{eq:lco}
\end{equation}
\ \\
\noindent with $\int_{\rm v} S_{\rm CO} \delta {\rm v} = I_{\rm CO}$ the integrated flux density of the CO(1-0) emission and $L'_{\rm CO}$ expressed in ${\rm K~km~s^{-1}~pc^2}$. 

Uncertainties in $L'_{\rm CO}$ are calculated following \citet{sag90}, assuming that they are dominated by the noise in the spectrum:
\begin{equation}
\Delta I_{\rm CO} = \sigma \Delta {\rm v} \sqrt{{{\rm FWZI}\over{\Delta {\rm v}}}}~(\rightarrow {\rm uncertainty~in}~L'_{\rm CO}) 
\label{eq:errorlco}
\end{equation}

\noindent with $\sigma$ the rms (root mean square) noise, $\Delta$\,v the channel width and FWZI the full width over which the CO profile was integrated. A second error term in $\Delta I_{\rm CO}$ arises from an uncertainty in determining the baseline ($\Delta I_{\rm CO}^{\rm baseline}$\,=\,$\sigma\Delta$${\rm v}_{\rm co}$$\sqrt{\Delta{\rm v_{\rm c}}/\Delta{\rm v_{\rm b}}}$, with $\Delta$v$_{\rm co}$ the FWZI of the CO signal, $\Delta$v$_{\rm c}$ the channel width and $\Delta$v$_{\rm b}$ the length of the baseline; see \citealt{sag90}). However, because of the large CABB bandwidth ($\Delta$\,v$_{\rm baseline}$\,$>>$\,$\Delta$\,v$_{\rm channel}$), this term is expected to be negligible. Our quoted uncertainties in $L'_{\rm CO}$ do not include the 20$\%$ uncertainty in absolute flux calibration (Sect.\ref{sec:observations}).

\begin{table*}
\caption{Observing details. $z$ and $\nu_{\rm CO(1-0)}$ are the redshift and central frequency at which we centred our `0'-velocity; t$_{\rm int}$ is the total on-source integration time; the last column lists the various calibrators that were used.}
\label{tab:obs}
\begin{tabular}{lcclcl}
\hline
\hline
Name & $z$  & $\nu_{\rm CO(1-0)}$ & Observing dates & t$_{\rm int}$ & Calibrators \\ 
 & &    (GHz) & & (h) & (P\,=\,phase, B\,=\,bandpass, F\,=\,flux) \\
\hline
MRC\,0114-211 & 1.402 & 47.998 & 03,09,22-AUG-11, 25,27-SEP-11, & 24.8 & 0130-171\,(P+B), 1921-293\,(B), \\
              & & &        23,24-OCT-11    &      &  Uranus\,(F)    \\   
MRC\,0152-209 & 1.9212 & 39.476 & 25,26-AUG-10, 28,29-SEP-10\,$^{\dagger}$ & 15.5 & 0130-171\,(P+B), Uranus\,(F) \\
MRC\,0156-252 & 2.016 & 38.220 & 05,07-AUG-10, 29-SEP-12 & 13.5 & 0135-247\,(P+B), Uranus\,(F)  \\
MRC\,0211-122 & 2.340 & 34.512 & 08,10-AUG-11, 26,27-SEP-11   & 10.9 & 0202-172\,(P+B), 0537-441\,(B), Uranus\,(F)    \\  
MRC\,0324-228 &  1.898  & 39.776 & 30-SEP-11, 20,21,22-OCT-11 & 11.1 & 0346-279\,(P), 0537-441\,(B), 1921-293\,(B), \\
              &  &  &                                 &      & 2223-052\,(B), Uranus\,(F)                     \\
MRC\,0350-279 & 1.900  & 39.749 & 30-SEP-11, 20,21,22-OCT-11 & 10.7 & 0346-279\,(P), 0537-441\,(B), 1921-293\,(B), \\
              &  &    &                               &       & 2223-052\,(B), Uranus\,(F)                \\
MRC\,0406-244 &  2.440  & 33.509 & 21,22-AUG-11, 15,16-SEP-11 & 14.0 & 0346-279\,(P), 0537-441\,(B), Uranus\,(F)  \\
MRC\,1017-220 & 1.768 & 41.644 & 24-OCT-11, 30,31-MAR-12, & 17.2 & 1034-293\,(P+B), Uranus\,(F) \\
              &   &        & 25,26,27-SEP-12          &  & Mars\,(F)    \\
MRC\,1138-262 & 2.161  & 36.467 & 18,19-AUG-11, 26,27-SEP-11 & 22.0 & 1124-186\,(P+B), Mars\,(F) \\
       & & & 13,14,15,16,17-MAR-12 & & \\
MRC\,2025-218  & 2.630  & 31.755 & 12,13,14,15-AUG-10 & 17.4 & 1958-179\,(P+B), 2008-159\,(P+B), 1934-638\,(F)  \\
MRC\,2048-272  &  2.060   &  37.670 &  21,22-AUG-11, 23-SEP-11 & 19.8 & 2058-297\,(P), 0537-411\,(B), 1921-293\,(B) \\
            &   & & 25-MAR-13, 03,04,05-OCT-13          &       & 2223-052\,(B), Uranus\,(F), G309\,(F) \\
MRC\,2104-242 & 2.491 & 33.020 & 18,19,20-SEP-09, 21,23-JUL-09 & 19.5 & 2008-159\,(P+B), 2128-123\,(P+B), 2149-306\,(P), \\
               &    & &        &  &  2223-052\,(B), Uranus\,(F)     \\ 
MRC\,2224-273  & 1.678  & 43.044 & 20,25,26-SEP-10, 01,02-OCT-10 & 15.3 & 2255-282\,(P+B), Uranus\,(F) \\
\hline
\hline
\end{tabular} 
\flushleft 
$^{\dagger}$ Additional high-resolution ATCA data of MRC\,0152-209 is presented in Emonts et al (in prep).
\end{table*} 

\begin{table*}
\caption{Observed CO(1-0) properties: `Beam' is the beam-size of the observations (with PA the position angle); rms is the root-mean-square noise in mJy\,beam$^{-1}$ per channel (no Hanning smooth applied); $\Delta$v is the channel width; FWZI is the full width at zero intensity of the CO(1-0) profile; $I_{\rm CO(1-0)}$ and $L'_{CO(1-0)}$ are the CO(1-0) intensity and luminosity; M$_{\rm H2}$ is the H$_{2}$ mass (see text for details); `$\sigma$-level' is the significance level of the CO(1-0) signal integrated across its entire velocity range (see text for details).}
\label{tab:results2}
\begin{tabular}{lcccccccc}
\hline
\hline
Name & beam (PA) & rms & $\Delta$v & FWZI & $I_{\rm CO(1-0)}$ & $L'_{\rm CO(1-0)}$ & M$_{\rm H2}$ & $\sigma$-level \\
MRC &  arcsec$^{2}$ ($^{\circ}$) & (mJy/bm) & (km/s) & (km/s) & (Jy/bm$\times$km/s) & (K\,\kms\,pc$^{2}$) & (M$_{\odot}$) & \\
\hline
0114-211 & 8.1\,$\times$\,6.0 (89) & 0.22 & 125 & 1130 & 0.43\,$\pm$\,0.08 & 4.5\,$\pm$\,0.9 $\times 10^{10}$ & $3.6\,\pm\,0.7 \times 10^{10}$ & 4.5$\sigma$ \\
0152-209 & 9.8\,$\times$\,7.1 (90)  & 0.23 & 40 & 800 & 0.37\,$\pm$\,0.04 & 6.6\,$\pm$\,0.8 $\times 10^{10}$ &  $5.3\,\pm\,0.6 \times 10^{10}$ & 9$\sigma$ \\
0156-252 &  6.9\,$\times$\,4.9 (71) & 0.17 & 154 & 2000 & 0.38\,$\pm$\,0.09 & 9.2\,$\pm$\,1.8 $\times 10^{10}$ & $7.4\,\pm\,1.5 \times 10^{10}$ & 5$\sigma$ \\
0211-122 & 11.1\,$\times$\,8.0 (99) & 0.19 & 91 & - & $<$\,0.17 & $<$\,$4.6 \times 10^{10}$ & $<$\,$3.7 \times 10^{10}$ & $<3\sigma$\\
0324-228 &  13.6\,$\times$\,10.6 (106) & 0.27 & 78 & - & $<$\,0.23 & $<$\,$4.2 \times 10^{10}$ & $<$\,$3.3 \times 10^{10}$  & $<3\sigma$\\
0350-279 & 13.7\,$\times$\,10.6 (102) & 0.28 & 78 & - &  $<$\,0.23 & $<$\,$4.3 \times 10^{10}$ & $<$\,$3.5 \times 10^{10}$  & $<3\sigma$\\ 
0406-244 & 12.4\,$\times$\,8.9 (77) & 0.14 & 95 & - &  $<$\,0.13 & $<$\,$3.7 \times 10^{10}$ & $<$\,$3.0 \times 10^{10}$  & $<3\sigma$\\
1017-220 & 6.7\,$\times$\,4.5 (84) & 0.20 & 74 & - & $<$\,0.16 & $<$\,$2.6 \times 10^{10}$ & $<$\,$2.1 \times 10^{10}$  & $<3\sigma$\\
1138-262 & 9.7\,$\times$\,6.1 (66)& 0.12 & 129 & 1680 & 0.31$\,\pm\,$0.06 & $7.2\,\pm\,1.3 \times 10^{10}$ & $5.8\,\pm\,1.0 \times 10^{10}$  & 8$\sigma$\\ 
2025-218 &  7.8\,$\times$\,4.8 (71) & 0.16 & 94 & - & $<$\,0.15 & $<$\,$4.8 \times 10^{10}$ & $<$\,$3.8 \times 10^{10}$  & $<3\sigma$\\
2048-272 & 7.1\,$\times$\,5.2 (89) & 0.10 & 210 & 3570 & 0.40\,$\pm$\,0.09 & $8.6 \pm 1.8 \times 10^{10}$ & $6.9 \pm 1.5 \times 10^{10}$  & 5$\sigma$\\
\,\,\,\,\,\,\,\,\,(A) &  &  &  & (1260) & ($0.21\,\pm\,0.05$) & ($4.5\,\pm\,1.1$) & ($3.6\,\pm\,0.9$) & \,\,\,(3.5$\sigma$) \\
\,\,\,\,\,\,\,\,\,(B) &  &  &  & (1260) & ($0.19\,\pm\,0.05$) & ($4.1\,\pm\,1.1$) & ($3.3\,\pm\,0.9$) & \,\,\,(3.5$\sigma$) \\
2104-242 & 11.7\,$\times$\,7.6 (97) & 0.15 & 101 & - & $<$\,0.14 & $<$\,$2.7 \times 10^{10}$ & $<$\,$2.1 \times 10^{10}$  & $<3\sigma$\\ 
2224-273 & 11.9\,$\times$\,10.0 (93) & 0.19 & 72 & - & $<$\,0.15 & $<$\,$2.2 \times 10^{10}$ & $<$\,$1.8 \times 10^{10}$  & $<3\sigma$ \\
\hline
\hline
\end{tabular} 
\flushleft 
\end{table*} 

\begin{figure*}
\centering
\includegraphics[width=\textwidth]{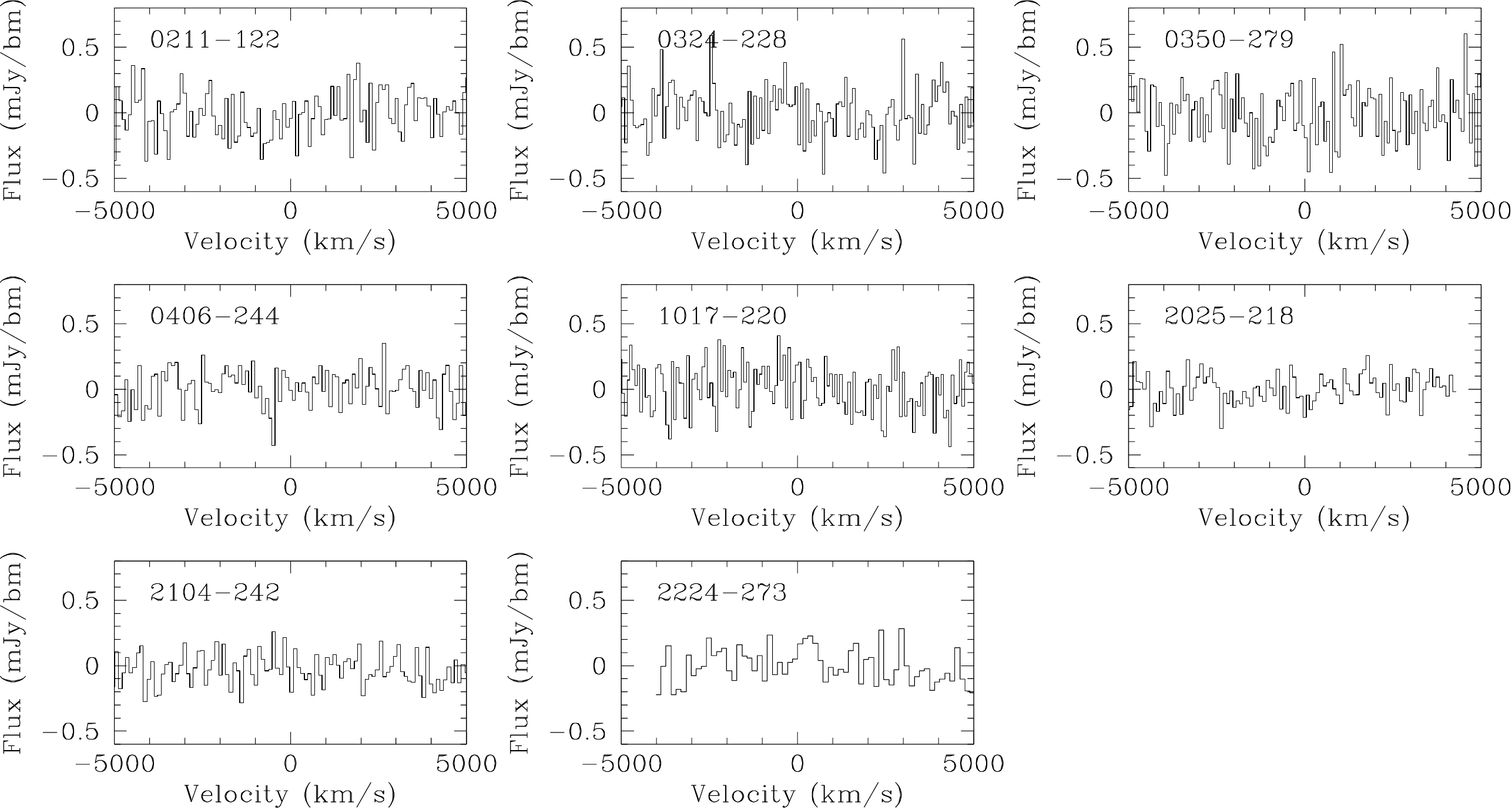}
\caption{Spectra of the sample sources that were not detection in CO(1-0). The spectra were binned by 10 channels to a velocity resolutions of roughly 100 \kms\ (see Sect\,\ref{sec:observations} and Table \ref{tab:results2}). The spectrum of MRC\,2224-273 (last window) was binned by 20 channels to visualise a low-level feature around the redshift of the object (which significance is too low to claim a detection).}
\label{fig:COupper}
\end{figure*}

To derive realistic upper limits for the non-detections in our sample, we set boundary conditions based on the CO(1-0) characteristics of our five CO-detected sources. They show CO(1-0) components with FWZI\,=\,$800-2000$\,\kms. In addition, the widest profiles (MRC\,0156-252, MRC\,1138-262 and MRC\,2048-272) appear to be double-peaked and cannot be fitted with a single Gaussian profile. This makes any assumption on the shape of the expected profile for the non-detections unjustified. We therefore derive conservative upper limits on $L'_{\rm CO}$ for the non-detected HzRGs in our sample by assuming a 3$\sigma$ signal smoothed across 1000\,\kms\ (FWZI):
\begin{equation}
I_{\rm CO} < 3\sigma \Delta {\rm v} \sqrt{{{\rm 1000}\over{\Delta {\rm v}}}}~~(\rightarrow {\rm upper~limits}~L'_{\rm CO}), 
\label{eq:errorlco}
\end{equation}
\noindent Results are given in Fig.\,\ref{fig:COupper} and Table \ref{tab:results2}.

\subsubsection{Significance level of the CO(1-0) detections}

An important consideration in the interpretation of our results is the significance of the CO(1-0) signals that we detect. In Table \ref{tab:results2} we show the signal-to-noise level of the {\sl integrated} CO(1-0) signal.\footnote{For MRC\,1138-262 (Fig.\,\ref{fig:mrc1138color}), MRC\,0152-209 (Fig.\,\ref{fig:mrc0152color}) and MRC\,0156-252 (Fig.\,\ref{fig:mrc0156lyaco}) the CO(1-0) signal could be separated into distinct kinematic components that peak at a different spatial location. In these cases, we derive the significance level of the `integrated' CO(1-0) profile as $\sqrt{\sum{(s_{c}^{\,\,2}}})$, with $s_{c}$ the $\sigma$-level of the different kinematic components. These values are somewhat higher than those shown in Fig.\,\ref{fig:COprofiles}, because one avoids adding noise from regions in which no CO(1-0) emission is found.} Most of the CO(1-0) detections presented in this paper are found $\la$1 synthesized beam-size from the core of the radio galaxy (Sect. \ref{sec:characteristics}) and near the redshift determined from optical emission-line observations \citep[][and references therein]{bre00}, resulting in a low probability that these signals are spurious \citep[see][]{kri12}. For sources that showed an initial tentative CO(1-0) signal (including MRC\,0114-211, MRC\,0156-252 and MRC\,2048-272), we continued observing them until the tentative CO(1-0) signal was either confirmed or judged spurious. The two main criteria that we used for confirming a CO(1-0) signal are: 1). indications for the CO(1-0) emission are present across the different observing runs; 2). at a channel-binning that is optimized to reveal the integrated CO(1-0) signal, the CO(1-0) emission stands out above the noise as the signal with the highest absolute value across the full primary beam and frequency coverage of the data cube. All five CO-detected sources in our sample satisfy these criteria. Appendix \ref{sec:app} provides additional details on our five CO(1-0) detections.

\citet{kri12} show that blind CO searches (i.e. searches for CO emission not associated with objects with known coordinates and redshifts) are likely to result in spurious CO detections at a 5$\sigma$ level \citep[see also][]{ara12}. Because many of the HzRGs in our sample have no, or no complete, information on proto-cluster galaxies in their environment, and because our observing strategy was not optimized for a blind CO(1-0) search, this paper does not address the CO(1-0) content of the larger (hundreds of kpc) environment of these HzRGs.

\subsection{CO characteristics}
\label{sec:characteristics}

In this Section we briefly discuss the characteristics of the CO(1-0) detections across our sample. Details on the individual sources are given in Appendix \ref{sec:app}.

\subsubsection{`On-source' CO(1-0) emission}
\label{sec:on}

\begin{figure*}
\centering
\includegraphics[width=\textwidth]{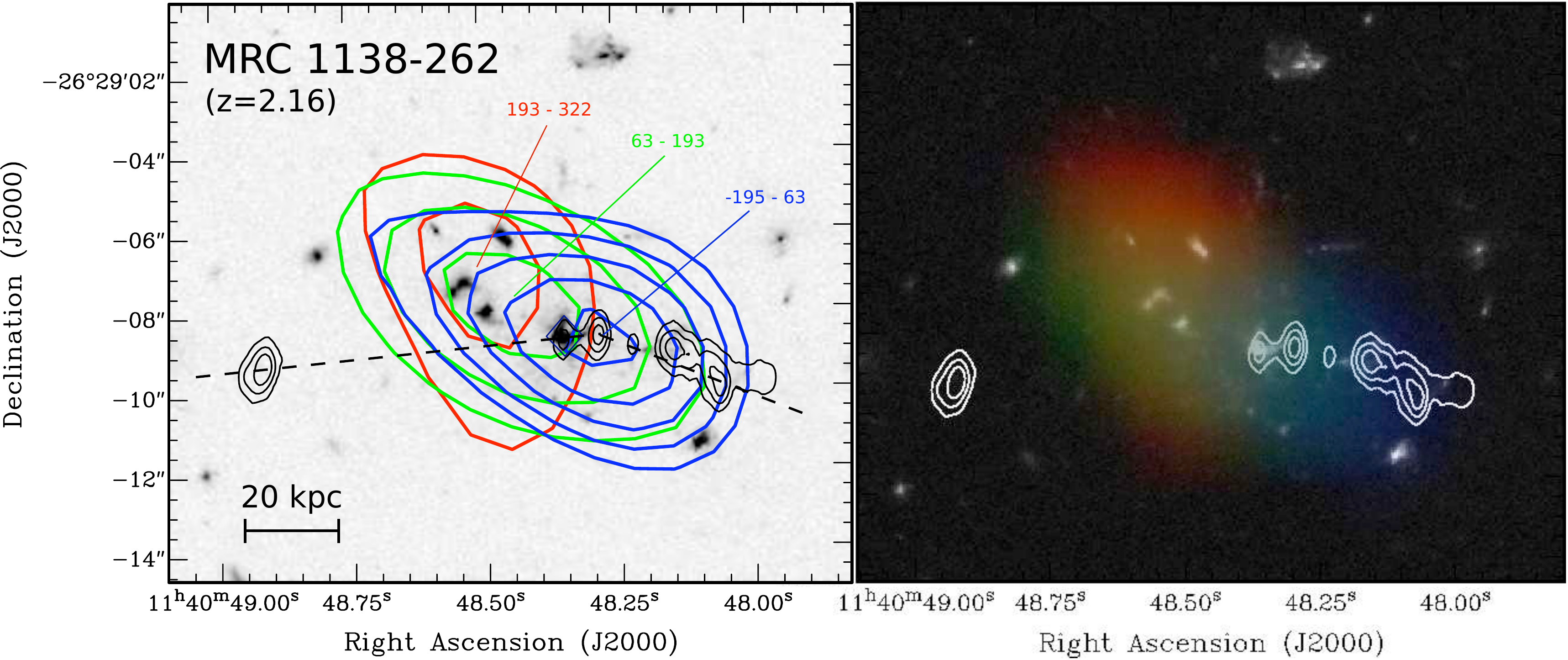}
\caption{Spatial distribution of the CO(1-0) emission associated with MRC\,1138-262 \citep[see also][]{emo13}. {\sl Left:} CO(1-0) emission integrated over three distinct velocity ranges and overlaid onto an {\sl HST/ACS\,(g$_{475}$+I$_{814}$)} image from \citet{mil06}. The velocity ranges for the three components are indicated in the plot (in \kms). A velocity gradient is visible in the CO(1-0) emission, which is spread across scales of 30-40 kpc (see \citealt{emo13} for details, including channel maps and spectra). CO(1-0) contour levels are at 3, 3.8, 4.6, 5.4, 6.2$\sigma$ level (with $\sigma$ = 0.033 Jy\,beam$^{-1}$\,$\times$\,\kms\ for the red/green components and 0.047 Jy\,beam$^{-1}$\,$\times$\,\kms\ for the blue component). The beam-size of the ATCA data is 9.7\('\)\('\)\,$\times$\,6.1\('\)\('\) (PA\,=\,66$^{\circ}$). The thin black contours show the 8.2\,GHz radio continuum image from \citet{car97}, while the black dotted line visualises the extrapolated radio jet axis. {\sl Right:} three-color image of the CO(1-0) components from the left plot overlaid onto the {\sl HST} image.}
\label{fig:mrc1138color}
\end{figure*}
\begin{figure*}
\centering
\includegraphics[width=\textwidth]{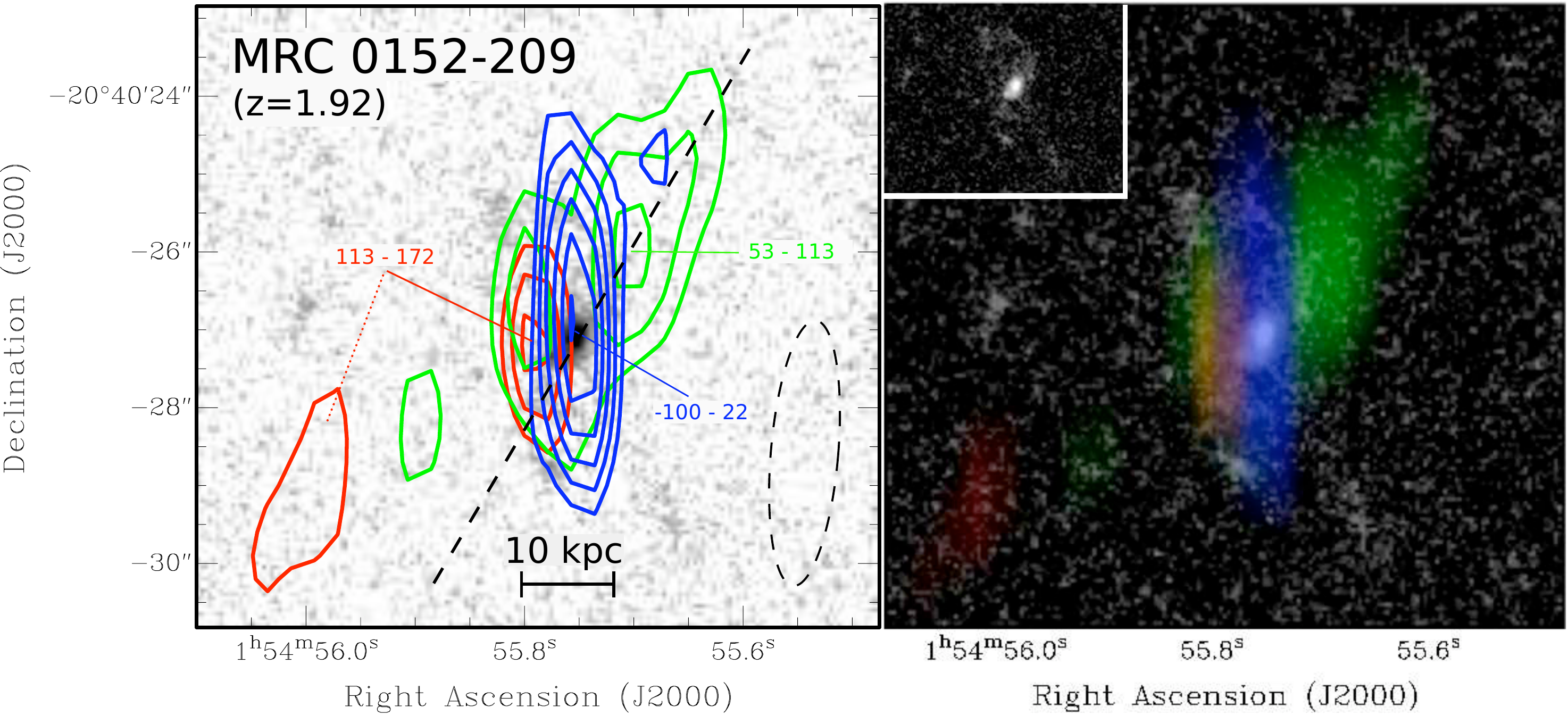}
\caption{Spatial distribution of the CO(1-0) emission associated with MRC\,0152-209 (see Emonts et al in prep for details, including channel maps and spectra; see \citealt{emo11a} for the original CO(1-0) detection). {\sl Left:} CO(1-0) emission integrated over three distinct velocity ranges and overlaid onto an {\sl HST/WFPC2}$_{\rm F555W}$ [Program 8183; PI: Miley] image (Emonts et al in prep). The velocity ranges for the three components are indicated in the plot (in \kms). CO(1-0) contour levels are at 2.8, 3.5, 4.2, 4.9, 5.6$\sigma$ level (with $\sigma$ = 0.0073 Jy\,beam$^{-1}$\,$\times$\,\kms\ for the red/green components and 0.012 Jy\,beam$^{-1}$\,$\times$\,\kms\ for the blue component). Note the highly elongated beam (dashed ellipse) that we obtained by using the extended-baseline East-West array configurations of the ATCA, which resulted in a lack of uv-coverage in North-South direction by staying above the typical 30$^{\circ}$ elevation suitable for millimetre observations (see Emonts et al in prep for details). The radio source (not shown in this plot) has a total linear extent of only 2.2 arcsec \citep[18 kpc;][]{pen00}, while the black dotted line visualises the extrapolated radio jet axis. {\sl Right:} three-color image of the CO(1-0) components from the left plot overlaid onto the {\sl HST} image. The inset in the top-left highlights the prominent optical tidal features.}
\label{fig:mrc0152color}
\end{figure*}

For two of the five CO(1-0) detected sources (MRC\,0152-209 and MRC\,1138-262) a substantial fraction of the CO(1-0) emission coincides with the bright optical/near-IR emission associated with the radio galaxy. In both cases, however, part of the CO(1-0) emission is also spread across scales of several tens of kpc (Figs.\,\ref{fig:mrc1138color} and \ref{fig:mrc0152color}). This is consistent with the idea that observations of the ground-transition CO(1-0) are sensitive to tracing wide-spread cold molecular gas \citep[][see also Sect,\,\ref{sec:intro}]{pap01}. Results on MRC\,0152-209 and MRC\,1138-262 are presented in detail in related papers \citep[][Emonts et al in prep.]{emo11a,emo13}, and will thus only be briefly summarised here. 

Figure \ref{fig:mrc1138color} shows the CO(1-0) distribution in MRC\,1138-262 \citep[from][]{emo13}.\footnote{Figure \ref{fig:mrc1138color} only includes the CO(1-0) features from \citet{emo13} that have recently been confirmed with additional ATCA observations (Emonts et al in prep). Several additional 3$\sigma$ features of the original data presented in \citet{emo13} have not yet been confirmed due to poor weather conditions during the new observations, hence they are not included in Fig.\,\ref{fig:mrc1138color}.} As discussed in \citet{emo13}, the kinematics of the CO(1-0) emission appears to follow the velocity distribution of several satellite galaxies in the same region, hence the extended part of the CO(1-0) is likely associated with (interacting/merging) satellites or the inter-galactic medium (IGM) between them.

\begin{figure}
\centering
\includegraphics[width=0.48\textwidth]{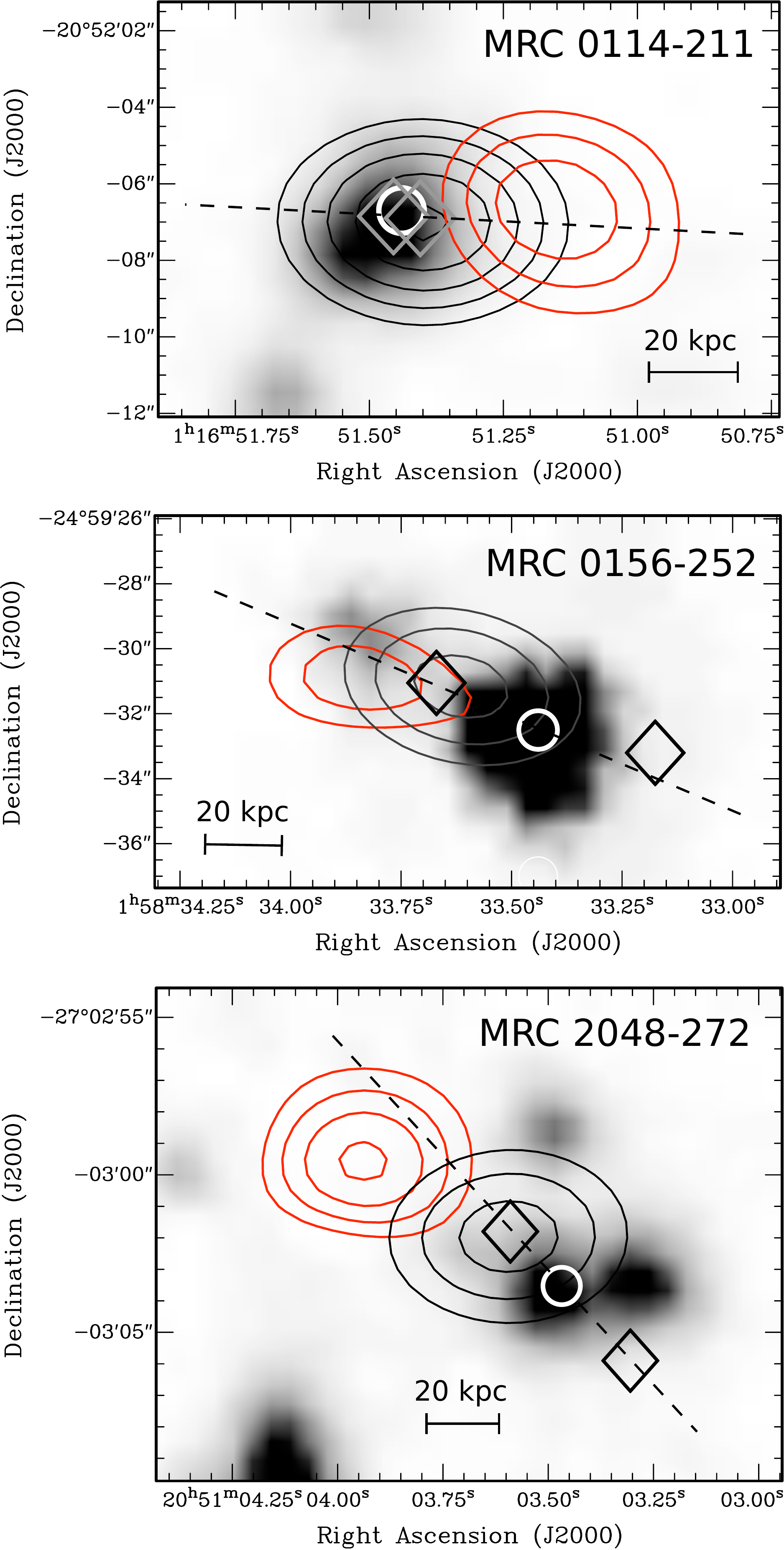}
\caption{Spatial distribution of the CO(1-0) emission associated with MRC\,0114-211, MRC\,0156-252 and MRC\,2048-272. The red contours show the CO(1-0) emission at 2.8, 3.5, 4.2, 4.9$\sigma$ level (with $\sigma$ = 0.094, 0.095 and 0.080 Jy\,beam$^{-1}$\,$\times$\,\kms\ for MRC\,0114-211, MRC\,0156-252 and MRC\,2048-272, respectively). The black contours represent the 7mm radio continuum (levels: 42, 52, 62, 72, 82 mJy\,beam$^{-1}$ for MRC\,0144-211; 6, 8, 10 mJy\,beam$^{-1}$ for MRC\,0156-252 and 1.6, 2.2, 2.8 mJy\,beam$^{-1}$ for MRC\,2048-272). The radio continuum and CO(1-0) emission were obtained from the same data set, so their relative astrometry is accurate. The lowest radio contour was chosen to represent the beam-size of the observations. The background plots are Spitzer/IRAC 4.5$\mu$m images from \citet{wyl13} and \citet{gal12}. The white circle shows the location of the radio galaxy from HST/NICMOS imaging \citep{pen01}, while the diamonds indicate the location of the radio hot-spots in high resolution radio imaging \citep[$\nu_{\rm obs} = 8$\,GHz;][]{pen01,bre10}. The dotted line visualises the extrapolated radio jet axis.}
\label{fig:COdistribution}
\end{figure}

Figure \ref{fig:mrc0152color} shows high-resolution ATCA data of MRC\,0152-209 (Emonts et al in prep), which is a follow-up study of our original low-resolution work \citep{emo11a}. MRC\,0152-209 is a gas-rich major merger system at $z=2$, with a double nucleus and prominent tidal tails in {\sl HST/NICMOS} and {\sl WFPC2} imaging \citep[][Emonts et al in prep]{pen01}. While the bulk of the CO(1-0) emission is co-spatial with the main body of the host galaxy, part of it seems to follow the optical tidal features. Interestingly, the small radio source appears to be aligned with the NW component of the off-nuclear CO(1-0) emission.
 
Galaxy mergers and interactions thus seem important to explain the CO(1-0) distribution in both MRC\,0152-209 and MRC\,1138-262. This is consistent with the suggestions by \citet{ivi12} that galaxy mergers are ubiquitous among starburst radio galaxies at high-$z$

\subsubsection{`Off-source' CO(1-0) emission}
\label{sec:off}

For the remaining three CO-detected sources in our sample (MRC\,0114-211, MRC\,0156-252 and MRC\,2048-272), the CO(1-0) emission is offset from the radio galaxy. As can be seen in Fig.\,\ref{fig:COdistribution}, for all three sources {\sl the CO(1-0) emission is located along the radio axis, on the side of the brightest radio emission, and found beyond the outer radio hot-spot.} The alignment of the CO with the radio jet axis is within 20$^{\circ}$ for the three sources. A fourth case may be the NW `off-source' component of CO(1-0) emission in MRC\,0152-209 (Sect.\,\ref{sec:on} and Fig.\,\ref{fig:mrc0152color}), as it shows a similar alignment with the small radio jet. These alignments resemble earlier results by \citet{kla04} and \citet{nes09}. It suggests that there is a causal connection between the propagating radio jets and the presence of large amounts of cold molecular gas in the halo environment of these sources (which we will discuss further in Sect.\,\ref{sec:alignment}). 

The peak of the CO(1-0) emission is located at an apparent $\sim$10, 30 and 40 kpc distance from the radio hot-spot for MRC\,0156-252, MRC\,0114-211 and MRC\,2048-272, respectively. However, the synthesized beam-size of our data corresponds to 60 - 70 kpc at these redshifts, and undetectable lower surface brightness radio plasma may be present beyond the hot-spot, so sensitive high resolution CO observations and deeper low-frequency radio imaging are required to further investigate how close the cold gas is located to the radio source.

\begin{figure}
\centering
\includegraphics[width=0.45\textwidth]{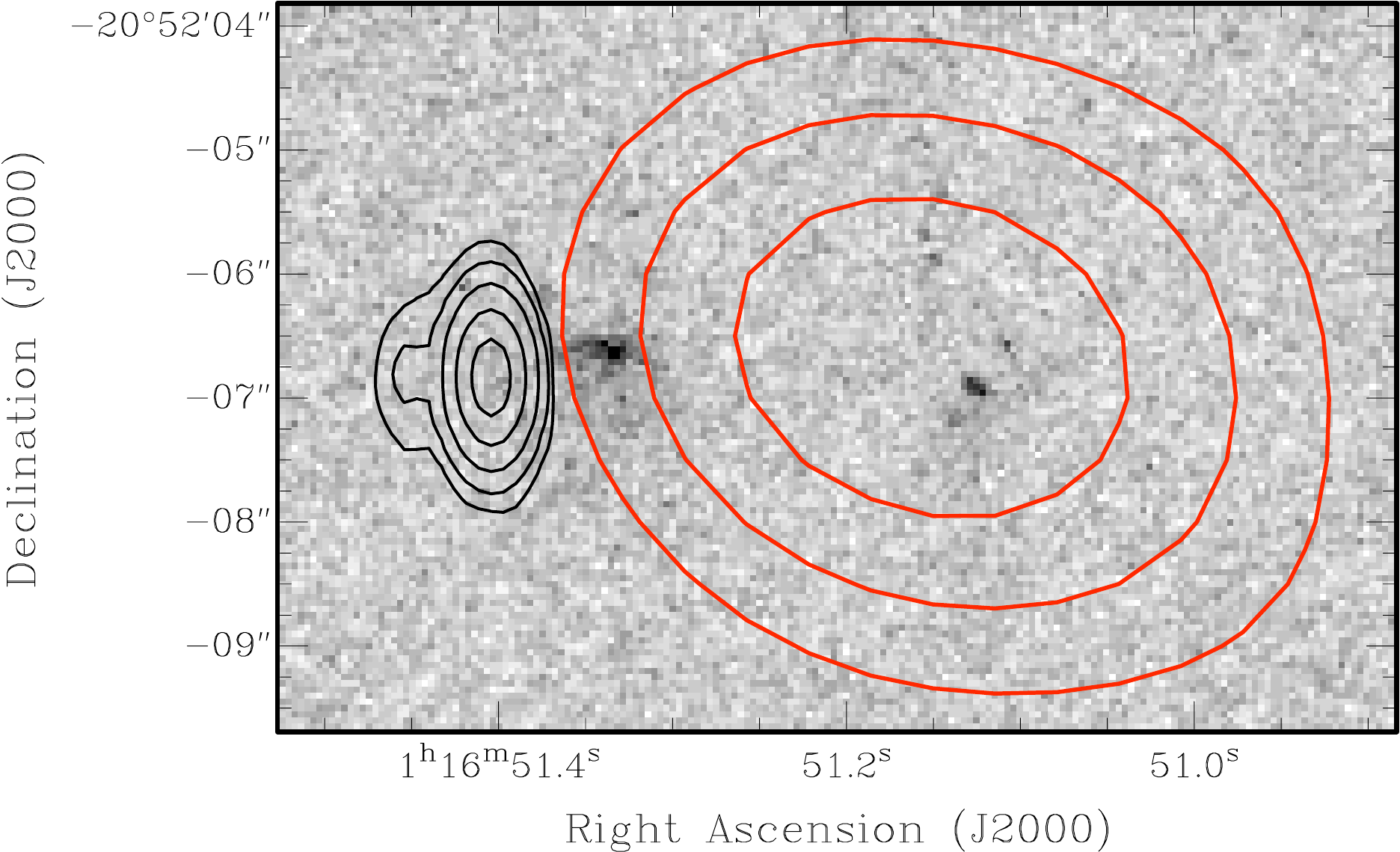}
\caption{HST/WFPC2$_{\rm F555W}$ image of the region around MRC\,0114-211, overlaid with the contours of CO(1-0) emission (red) and 4.8\,GHz radio continuum from \citet{bre10} (black). Contour levels CO(1-0): 2.8, 3.5, 4.2$\sigma$ (with $\sigma$ = 0.094 Jy\,beam$^{-1}$\,$\times$\,\kms). The accuracy of the HST astrometry has not been confirmed to better than $\sim$1.5\,arcsec (see text).}
\label{fig:app0114}
\end{figure}

While $L'_{\rm CO}$ of these three off-nuclear CO-emitters is similar to what is found in submillimetre galaxies (SMGs; see Sect.\,\ref{sec:samples}), Fig. \ref{fig:COdistribution} shows that no {\sl Spitzer/IRAC} 4.5$\mu$m emission is found at the location of the CO(1-0) to a level of one to two magnitudes below $L^{*}$ \citep[see][]{gal12,wyl13}. For MRC\,0114-211, however, the CO(1-0) emission is co-spatial with a very faint optical (i.e. near-UV rest-frame) counterpart that is visible in an archival {\sl HST/WFPC2}$_{\rm F555W}$ image (Fig.\,\ref{fig:app0114}). Brighter optical emission is found just outside the edge of the bright western radio jet, reminiscent of the shell of shocked emission-line gas, such as the one that \citet{ove05} found around the bended radio lobe in MRC\,0156-252. It is possible, however, that the error in the astrometry of the archival {\sl HST} image is as much as 1.5\,arcsec (see Appendix\,\ref{sec:app}) and that the bright optical emission is the host galaxy.

\begin{figure}
\centering
\includegraphics[width=0.47\textwidth]{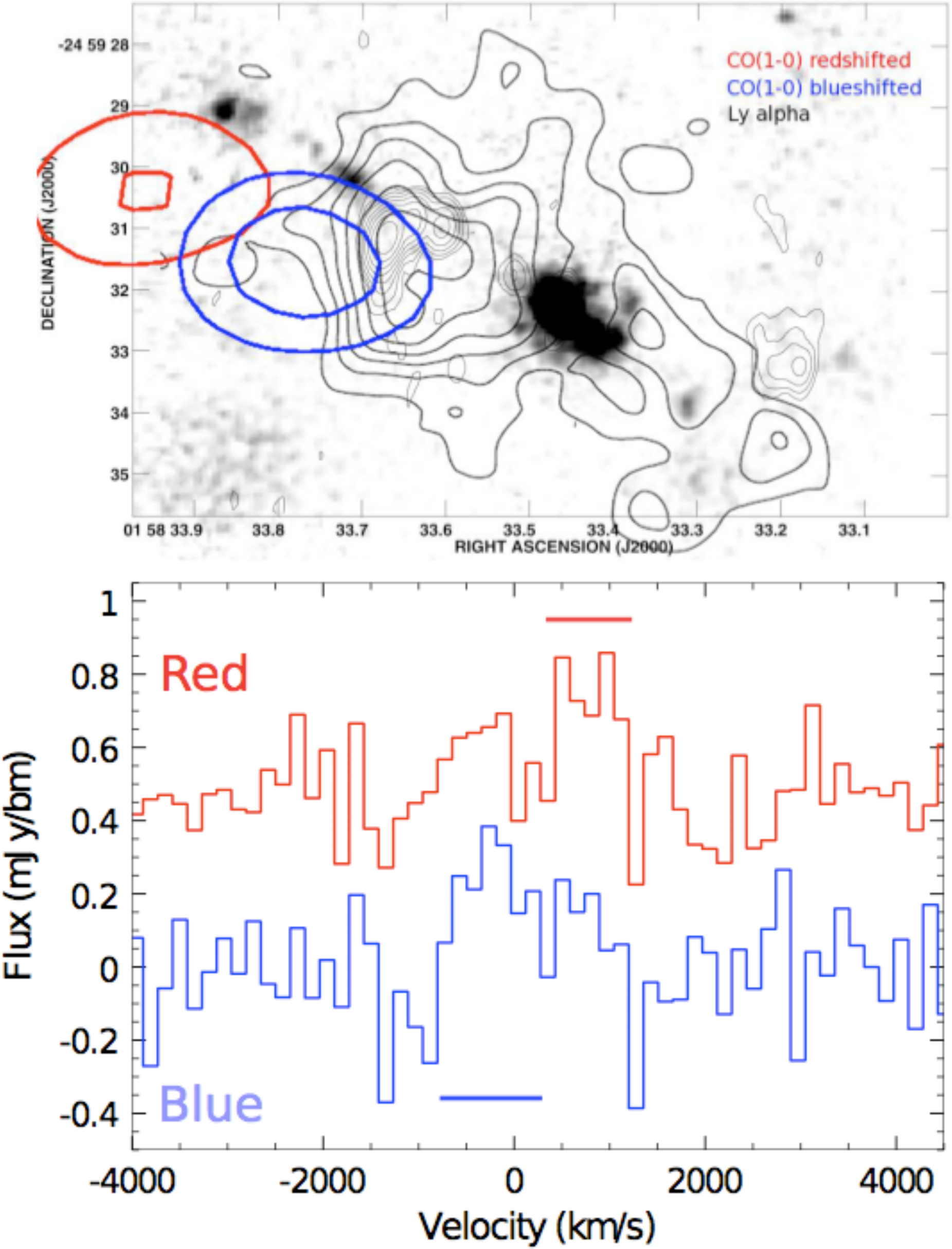}
\caption{{\sl Top:} Spatial distribution of the CO(1-0) emission associated with MRC\,0156-252. The blue and red contours are the blue- and redshifted part of the CO(1-0) profile from Fig.\,\ref{fig:COprofiles} (integrated over the velocity ranges -800\,$-$\,+275 and +275\,$-$\,+1200 \kms, respectively, as indicated by the bars in the bottom plot). Contour levels are at 2.8 and 3.5 $\sigma$ level (with $\sigma = 0.055$ and 0.064 Jy\,beam$^{-1}$\,$\times$\,\kms\ for the red and blue component, respectively). The thick black contours are \lya\ emission, the thin black contours 8.2 GHz radio continuum and the grey-scale HST/NICMOS imaging, all from \citet{pen01}. {\sl Bottom:} Spectra at the location of the blue and red peak in the top plot. As can be seen, both spectra are not fully independent due to the large beam-size of our observations.}
\label{fig:mrc0156lyaco}
\end{figure}

Fig.\,\ref{fig:mrc0156lyaco} shows that the CO(1-0) emission in MRC\,0156-252 is located just outside a large reservoir of \lya\ emission, with the boundary region between the \lya\ and CO(1-0) gas occurring at the edge of the radio hot-spot. The CO(1-0) emission consists of two kinematically distinct peaks that both lie $\sim$20 kpc distance from two satellite galaxies detected with {\sl HST/NICMOS} \citep{pen01}. The innermost companion (near the blue CO peak) has blue colors that are indicative of enhanced star formation \citep[possibly triggered by the passage of the radio jet;][]{pen01}. Very recent work by \citet{gal13} suggests that the redshift at which we centred the CO(1-0) profile ($z=2.016$) is more closely related to the redshift of this blue companion than that of the HzRG (see Appendix \ref{sec:app} for details).

For MRC\,2048-272, the two kinematically distinct CO(1-0) components from Fig.\,\ref{fig:COprofiles} peak at the same spatial location (see Appendix\,\ref{sec:app} for details).

We also note that the CO(1-0) emission of these three off-nuclear detections covers a wide velocity range (1100\,$<$\,FWZI\,$<$\,3600\,\kms). The CO(1-0) profiles are broad compared to what is generally found in quasars and submillimetre galaxies \citep[][]{cop08,wan10,ivi11,rie11,bot12,kri12}, with the exception of what is found in a few high-$z$ merging galaxies \citep[see][and references therein]{sal12}. Possible scenarios on the nature of the off-nuclear CO(1-0) emission in MRC\,0114-211, MRC\,0156-252 and MRC\,2048-272 are discussed in Sect.\,\ref{sec:alignment}

\subsection{Derivation of H$_{2}$ mass}
\label{sec:H2mass}

Molecular gas masses (and upper limits) are estimated from $L'_{\rm CO}$ using a conversion factor $X_{\rm CO}$ = M$_{\rm H2}$/$L'_{\rm CO}$ = 0.8 M$_{\odot}$ (K \kms\ pc$^{2}$)$^{-1}$ \citep[where M$_{\rm H2}$ includes a helium fraction; e.g.][]{sol05}. This value is found for Ultra Luminous Infra-Red Galaxies \citep[$L_{\rm IR} > 10^{12} L_{\odot}$;][]{dow98} and is within the range of $X_{\rm CO}$\,=\,$0.8-1.6$ M$_{\odot}$ (K \kms\ pc$^{2}$)$^{-1}$ assumed for high-$z$ submillimetre galaxies and star forming galaxies \citep{tac08,sta08}. However, we stress that $X_{\rm CO}$ depends on important properties of the gas \citep[such as metallicity, extinction and radiation field;][]{pap08,glo11,bol13}, and is not yet well understood \citep[][see additional discussion in Sect. \ref{sec:enrichment}]{tac08,ivi11}. 

Taking the above into account, our assumed $X_{\rm CO}$\,=\,0.8 M$_{\odot}$ (K \kms\ pc$^{2}$)$^{-1}$ results in molecular gas mass estimates in the range M$_{\rm H2} = 4-7 \times 10^{10}$\,M$_{\odot}$ for the CO(1-0) detected HzRGs in our sample. Value and upper limits of M$_{\rm H2}$ are given in Table \ref{tab:results2}.

\section{Discussion}
\label{sec:discussion}

We have performed a systematic search for CO(1-0) emission of cold molecular gas in a representative sample of 13 high-$z$ radio galaxies and have presented five CO(1-0) detections. The most intriguing result from this work is that we find three cases where bright CO(1-0) emission is found along the radio jet axis and beyond the outer radio hot-spot. This indicates that the alignments seen in HzRG between the radio jets and optical/UV \citep{mcc87,cha87}, X-ray \citep{car02,sma13} and submillimetre emission \citep{ste03} now also have to take into account the component of cold molecular CO(1-0) gas. 

In Sect.\,\ref{sec:cofir} we will first present a comparison between $L'_{\rm CO(1-0)}$ and the far-IR luminosity ($L_{\rm FIR}$) from the starburst component. We will use this information in Sect. \ref{sec:alignment} to interpret the alignments that we find between the radio source and cold molecular halo gas in several of our sample sources. A more statistical investigation of the radio\,-\,CO alignment, based on CO results from the literature, is given in Sect.\,\ref{sec:statistics}. In Sect.\,\ref{sec:samples} we will compare the CO(1-0) properties of HzRG with those of other types of high-$z$ galaxies.

\subsection{Cold gas ($L'_{\rm CO}$) vs. starburst ($L_{\rm FIR}$)}
\label{sec:cofir}
 
\begin{figure}
\centering
\includegraphics[width=0.47\textwidth]{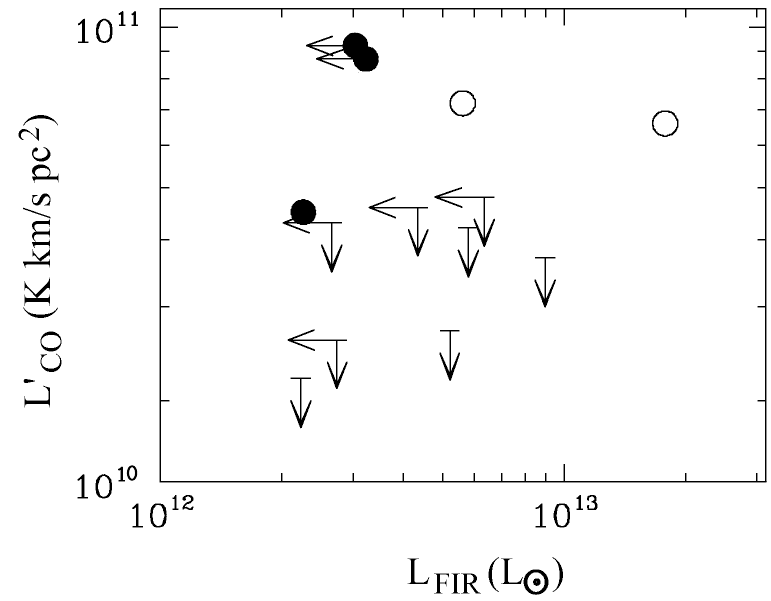}
\caption{$L'_{\rm CO}$ plotted against the starburst far-IR luminosity $L_{\rm FIR}$ ($8-1000 \mu$m). Values of $L_{\rm FIR}$ are from Drouart et al (in prep). The open circles represent MRC\,0152-209 and MRC\,1138-262, in which a large fraction of the CO(1-0) emission coincides with the central radio galaxy. The filled circles show the three sources where the CO(1-0) emission is solely detected offset from the radio galaxy.}
\label{fig:fir}
\end{figure}

Following the Schmidt-Kennicutt relation between star formation and gas reservoir \citep{sch59,ken98}, relations between $L'_{\rm CO}$ and the far-IR luminosity ($L_{\rm FIR}$) are frequently observed in both low- and high-$z$ galaxies \citep[see][for recent examples]{ivi11,vil13}. Figure \ref{fig:fir} shows $L'_{\rm CO}$ plotted against the starburst component of $L_{\rm FIR}$ for our sample sources. Values of $L_{\rm FIR - starburst}$ (hereafter $L_{\rm FIR}$) are from Drouart et al (in prep) and have been derived from modeling the Spectral Energy Distribution (SED) across $8-1000\,\mu$m to separate the starburst from other, e.g. torus and stellar, components. Details and a list of $L_{\rm FIR}$ values will be provided in Drouart et al (in prep). 

Given the small range of $L'_{\rm CO}$ and $L_{\rm FIR}$ values that our sample covers, it is not surprising that there is no clear trend visible. However, it is interesting that the sources in which we detect CO(1-0) solely offset from the central host galaxy (MRC\,0114-211, MRC\,0156-252 and MRC\,2048-272) have a starburst FIR luminosity that falls in the lower half of the $L_{\rm FIR}$ values of our sample sources. This suggests that earlier studies that pre-selected targets based on their high FIR luminosity may have missed similar systems where detectable amounts of molecular gas are located outside the host galaxy, but in the galaxy's halo environment.

MRC\,1138-262 and MRC\,0152-209 (the two sample sources where part of the CO(1-0) emission coincides with the host galaxy) are among the stronger FIR emitters in the sample. Their high FIR luminosity implies high star formation rates \citep[up to $\sim$1400 M$_{\odot}$\,yr$^{-1}$;][]{emo11a,sey12}. As discussed in \citet{emo11a,emo13}, if the CO(1-0) represents a reservoir of molecular gas that is consumed by this massive star formation, the minimum gas depletion time-scale in these two HzRGs is of order 40 Myr. This is comparable to the typical life-time of a massive burst of star formation in ultra-luminous IR ($L_{\rm IR} > 10^{12} L_{\odot}$) merger systems \citep[e.g.][]{mih94}.

When comparing the CO(1-0) luminosity of our sample sources with other properties of the host galaxy \citep[as observed by][]{bre10, pen00radio}, we find no apparent correlation between $L'_{\rm CO}$ and $z$, P$_{\rm 500\,W\,Hz^{-1}}$, spectral index $\alpha^{\rm 4.7\,GHz}_{\rm 8.2\,GHz}$, total linear extent of the radio source and core fraction of 20\,GHz radio continuum. Three of our CO(1-0) detected sources (MRC\,1138-262, MRC\,0156-252 and MRC\,0152-209) appear to have a relatively high stellar mass (in the range $0.5 - 2 \times 10^{12}$\,M$_{*}$), compared to the CO non-detections \citep[M$_{*} < 0.5 \times 10^{12}$\,M$_{\odot}$;][]{bre10}. With better statistics it would be worth investigating whether there is a trend between $L'_{\rm CO(1-0)}$ and M$_{*}$.

\subsection{Cold halo gas: radio - CO alignment}
\label{sec:alignment}

Our survey suggests that CO reservoirs of cold molecular gas exist in the halo ($\sim$50\,kpc-scale) environment of a significant fraction of HzRGs. These cold gas reservoirs are likely part of metal-enriched quiescent \lya\ halos that have been observed to surround HzRGs \citep{vil03,vil06,vil07,bin06}. Extended reservoirs of neutral hydrogen gas and large dusty shells have also been observed in these \lya\ halos \citep{oji97lya,jar03,hum08,hum13}.

The most intriguing result from this paper is that this CO(1-0) is preferentially aligned along the radio jet axis and found beyond the brightest part of the radio continuum. This is similar to the case of TXS\,0828+193 ($z=2.6$), where \citet{nes09} found CO(3-2) emission ($L'_{\rm CO} \sim 2 \times 10^{10}$ $\times$ ${\rm K~km~s^{-1}~pc^2}$) beyond the tip of the SW hot-spot ($\sim$80 kpc from the radio core). No optical or IR counterpart was detected at this location and the CO kinematics are in good agreement with those of C\,{\small IV} emission at the edge of the optical gaseous halo \citep{vil02}. Our results also resemble the ${\rm radio} - {\rm CO}$ alignments found in $z>3$ HzRGs by \citet{kla04} as well as in the $z \sim 0.3$ quasars [H89]1821+643 \citep[][]{blu01,ara11} and HE\,0450-2958 \citep[][]{kla07,pap08,elb09}. We here address several possible explanations for this alignment.

\subsubsection{Jet-induced star formation or gas cooling}

\citet{kla04} found that in $z > 3$ radio galaxies CO and dust emission are also preferentially aligned along the radio axis. They discuss a scenario in which the CO is formed in sites of star formation that are initially triggered by the radio jets -- a mechanism that may perhaps also explain alignments found between the radio jets and UV/optical continuum \citep{mcc87,cha87,ree89,you89,beg89,bic00}, or the increased star formation rates found in AGN with pronounced radio jets \citep{zin13}. \citet{kla04} make three predictions for the scenario of jet-induced star formation that our independent CO(1-0) results appear to confirm, namely that the CO in HzRGs is extended and aligned with observable synchrotron radio emission (see our Fig.\,\ref{fig:COdistribution}), that the CO profiles from emission outside the host galaxy will be broad (see our Fig.\,\ref{fig:COprofiles}) and that CO and \lya\ emission will be tracing different physical regions (see our Fig.\,\ref{fig:mrc0156lyaco}).

It is possible that at $z \sim 2$ (where the massive radio galaxies have already gone through several cycles of chemical enrichment) the CO is associated with jet-induced gas cooling that precedes the star formation, rather than being a product of the star formation itself. Gas cooling is feasible if the IGM is compressed and cooled through shocks induced by the radio source \citep[e.g.][]{mel02,sut03,fra04,gai12}. As discussed in detail by \citet{nes09}, this may also result in cooling-flow processes similar to those observed in low-$z$ giant central cluster galaxies \citep[despite morphological and kinematical differences in the CO distribution, which in low-$z$ cooling flows is mainly concentrated towards the central galaxy and along the sides of the cavities that are excavated by the radio lobes; e.g.][]{fab94,edg01,sal03,sal06,sal11}.

\subsubsection{Jet-driven enrichment}
\label{sec:enrichment}

An alternative scenario is that the carbon and oxygen were created during massive bursts of star formation within the host galaxy and subsequently transported into the halo environment by the radio source, forming a reservoir of metal-enriched gas beyond the edge of the radio plasma. 


Jet-driven outflows of ionised, neutral and molecular gas have been observed along the radio axis in both low- and high-z radio galaxies \citep[e.g.][]{vil03,mor03,mor05,emo05,hum06,nes08,hol08,das12,mor13a,mor13b,mah13}. In addition, it is believed that powerful radio-loud AGN are responsible for the formation of filamentary Extended Emission Line Regions (EELRs) in quasars \citep[][]{fu07a,fu07b,fu09}. Several studies suggest that the radio jets may also be responsible for transporting dust and metals far outside the host galaxy. In low-$z$ brightest clusters galaxies, \citet{kir09,kir11} find that metal-enrichment takes place along the radio axis. They show that heavy elements are often found beyond the extent of the innermost X-ray cavities, which suggests that the metal-enrichment by the radio source is sustained over multiple generations of radio outbursts. At high-$z$, \citet{ivi12} argue that jet-induced feedback may explain a large ($\sim500$\,kpc) dusty filamentary structure that is co-aligned with the radio source 6C\,1909+72 ($z=3.5$).


Interestingly, extended emission-line halos around HzRGs and quasars have been found to display near-solar metallicities \citep[e.g.][]{ver01,hum08shocks,pro09}. It is interesting to speculate that in this metal-rich halo environment, the potentially low gas densities and high velocity dispersion of the cold gas (as traced by the wide FWZI of our CO profiles), in combination with a potentially high-pressure environment induced by the radio plasma and the absence of a strong stellar radiation field, may conspire in favor of a lower conversion factor than our assumed $X_{\rm CO} = 0.8$ \citep[see][]{pap08,glo11}. This would lower the mass of molecular gas responsible for the bright CO(1-0) emission and alleviate the problem that this cold halo gas is found in regions devoid of {\sl Spitzer} 4.5$\mu$m emission. Because the CO(1-0) emission in the halo is produced in a profoundly different environment than models on the $X_{\rm CO}$ factor generally assume, a more detailed analysis of this is beyond the scope of this paper.

When considering the timescales involved, assuming that the enriched material is expelled with a typical velocity of $\sim$500 \kms \citep[][]{nes08}, it would take the gas $\sim 6 \times 10^{7}$ yr to reach a distance of 30 kpc. In agreement, the typical lifetime of extended radio sources is expected to be at least several $\times 10^{7-8}$ yrs \citep[][]{par99,blu00}. \citet{ove05} argue that the nuclear activity in MRC\,2048-272 has recently ceased (given the lack of X-ray and radio continuum in the core), indicating that this system is at the end of its current radio-loud AGN cycle.

In addition, in several $\times$ 10$^{7}$ yr a major starburst episode may have passed its peak activity \citep[e.g.][]{mih94}, which may perhaps explain why the hosts of the off-nuclear CO(1-0) emitters are fainter in the FIR than MRC\,0152-209 and MRC\,1138-262, which have the bulk of CO (still) centred on the host galaxy. While for MRC\,0152-209 part of the CO(1-0) appears to extend along the radio axis (Sect.\,\ref{sec:on}), it is also interesting to note that for MRC\,1138-262 \citet{nes06} detect a fast, redshifted ionised gas outflow along the same direction as the extended CO(1-0) emission (see \citealt{emo13} for a discussion).

If jet-driven enrichment is a viable scenario for the radio - CO alignment, we can make the testable predictions that {\sl (i)} the chemical enrichment of the halos around HzRGs should occur mainly along the radio jet axis, {\sl (ii)} HzRGs with CO reservoirs in their halos may have passed a peak-period of major starburst activity and {\sl (iii)} there should be high-$z$ galaxies in which the radio source has recently switched off (e.g. radio-quiet QSOs) with similar off-nuclear CO(1-0) reservoirs.


\subsubsection{Jet-brightness enhancement by cold ISM}
\label{sec:alternative}

Alternatively, the CO(1-0) emission may represent tidal debris from the host or companion galaxies (similar to the tidal CO(1-0) gas we find around MRC\,0152-209; Fig.\,\ref{fig:mrc0152color}), or a filamentary structure of cold molecular gas \citep[possibly an imprint of the large-scale Cosmic Web in the center of the proto-cluster;][]{wes91,spr06,cev10}. In those cases, the radio - CO alignment effect could occur because the working surface of the radio jet is brightest there where it encounters the densest medium \citep[e.g.][]{bar96}. The characteristic bright radio continuum in HzRGs would then be a consequence of the fact that the synchrotron jets propagate into a high density region. This scenario was also used by \citet{ste03} to explain their observed alignment between the radio source and extended submillimetre emission in and around HzRGs (and has also been invoked to explain the relatively high occurrence of bright, compact radio sources in gas-rich and starbursting radio galaxies at low-$z$; e.g. \citealt{tad11,mor11,emo12}).

\subsection{Radio - CO alignment: general statistics}
\label{sec:statistics}

\begin{figure*}
\centering
\includegraphics[width=\textwidth]{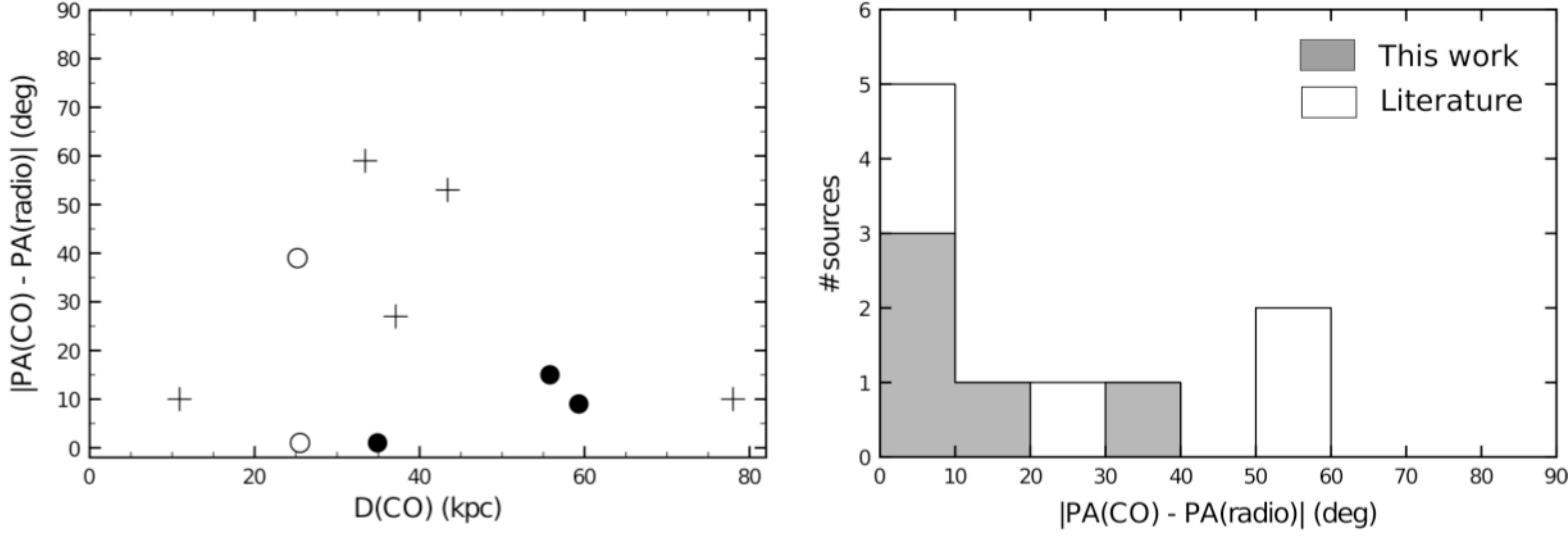}
\caption{{\sl Left:} Difference in position angle ($\psi$) between the CO and synchrotron radio emission in HzRG (Table\,\ref{tab:alignment}) plotted against the distance to which the CO emission extends. The filled and open circles are the same as in Fig.\,\ref{fig:fir} and represent our sample sources detected in CO(1-0). The crosses are HzRGs with extended CO($J,J-1$) emission from the literature. {\sl Right:} Histogram of the distribution of $\psi$ among HzRGs with extended CO($J,J-1$) emission.}
\label{fig:alignment}
\end{figure*}

\begin{table*}
\caption{Literature study on the potential alignment of CO($J,J-1$) emission with radio source axis in HzRGs. PA$_{\rm CO}$ and PA$_{\rm R}$ are the position angle of the CO emission and the radio jet (on the side where the CO emission occurs). $\psi$ = $|$PA$_{\rm CO}$ - PA$_{\rm R}$$|$ is the difference in position angle between the CO and radio jet. D$_{\rm CO}$ and D$_{\rm R}$ are the maximum distance out to which the CO and radio emission are detected. R = D$_{\rm CO}$/D$_{\rm R}$ is the ratio of these values, which thus gives an indication how far inside (R\,$<$\,1) or outside (R\,$>$\,1) the radio jet the CO emission is found. References: 1. \citet{bre05}; 2. \citet{car94}; 3. \citet{pap00}; 4. \citet{ivi08}; 5. \citet{car97}; 6. \citet{pen01}; 7. \citet{bre03AR}; 8. \citet{ivi12}; 9. \citet{per05}; 10. \citet{nes09}; 11. this work}
\label{tab:alignment}
\begin{tabular}{lccccccccc}
\hline
\hline
Name & $z$  & CO & PA$_{\rm CO}$ & PA$_{\rm R}$ & $\psi$ & D$_{\rm CO}$ & D$_{\rm R}$ & R\,(D$_{\rm CO}$/D$_{\rm R}$) & Refs. \\ 
     &      &    transition    & ($^{\circ}$)  & ($^{\circ}$)  & ($^{\circ}$) & (kpc) & (kpc)  & & \\
\hline
4C\,41.17      & 3.792 & (4-3) & -140 & -130 & 10 & 10.9 & 57.9 & 0.19 & 1,2 \\
4C\,60.07      & 3.791 & (4-3) & -70 & -123 & 53 & 43.4 & 45.6 & 0.95 & 3,4,5 \\
6C\,1908+7220    & 3.537 & (4-3) & -127 & -154 & 27 & 37.1 & 50.5 & 0.73 & 6,7 \\
B3\,J2330+3927 & 3.086 & (1-0)/(4-3) & -154 & -213 & 59 & 33.4 & 15.5 & 2.15 & 8,9 \\
TXS\,0828+193  & 2.6   & (3-2) & -131 & -141 & 10 & 78.0 & 56.9 & 1.37 & 2,10 \\
MRC\,1138-262 & 2.161 & (1-0) & 56 & 95 & 39 & 25.2 & 63.8 & 0.39 & 11 \\
MRC\,0156-252 & 2.016 & (1-0) & 76 & 67 & 9 & 59.3 & 25.4 & 2.33 & 11 \\
MRC\,2048-272 & 2.060 & (1-0) & 58 & 43 & 15 & 55.8 & 22.8 & 2.45 & 11 \\
MRC\,0152-209 & 1.921 & (1-0) & -31 & -32 & 1 & 25.5 & 18 & 1.42 & 11 \\
MRC\,0114-211  & 1.402 & (1-0) & -92 & -93 & 1 & 34.9 & 4.3 & 8.12 & 11 \\
\hline
\hline
\end{tabular} 
\flushleft 
\end{table*} 

As mentioned in Sect.\,\ref{sec:characteristics}, for all three HzRGs with CO(1-0) emission found solely at large distance from the host galaxy, the CO(1-0) emission is aligned to within 20$^{\circ}$ of the radio axis. For MRC\,0152-209, the off-nuclear the CO(1-0) emission towards the NE is also appears to be aligned with the small radio jet (Sect. \ref{sec:on}). The chance of CO(1-0) being aligned within 20$^{\circ}$ for four out of the five CO-detected sources in our sample is $\la 1\%$. However, the beam-size of our observations is large and these CO(1-0) signals peak at only $\sim$5$\sigma$ significance, introducing substantial uncertainty in the exact location of the CO(1-0) emission. Moreover, given the low number of CO-detected sources in our sample, we cannot place reliable statistical significance on these results.

In order to investigate in a more statistical way whether there is evidence for a general alignment between CO emission and radio jets in HzRGs, we show in Table \ref{tab:alignment} and Fig.\,\ref{fig:alignment} a summary of all HzRGs (with $L_{500} > 10^{27}$ W\,Hz$^{-1}$) from the literature for which extended CO and radio emission have been imaged. We use a one-tailed Mann-Whitney U-test to verify the hypothesis that the CO emission in this sample is preferentially aligned towards the radio jet axis, compared to samples of equal size where the differences in position angle between the CO and radio are presented by randomly chosen numbers between 0 and 90 degrees. We find marginal statistical significance at the 95$\%$ level that such an alignment is present in the data.

These results are a largely independent confirmation of the results by \citet{kla04}, who derived their conclusions from a sample of high-$z$ radio sources with a lower average radio power. We note, however, that we are still dealing with low-number statistics and that most of the studies from the literature where done with higher order CO transitions (which likely trace a different component of the gas than CO(1-0); Sect.\,\ref{sec:intro}). Larger samples observed with ALMA and the EVLA are required to verify the alignment effect for various $J$-transitions of CO.

\subsection{The CO(1-0) content of HzRGs}
\label{sec:samples}

We detect CO(1-0) emission (with $L'_{\rm CO} > 4 \times 10^{10}$\,L$_{\odot}$) associated with 38$\%$ of our sample of southern HzRGs. In earlier studies \citet{oji97} did not detect any CO among a sample of 14 northern HzRGs \citep[see also][]{eva96}. \citet{oji97} derived upper limits of L'$_{\rm CO} \sim {\rm few} \times 10^{10}$ K\,\kms\,pc$^{2}$, but mainly using the higher order $J$-transitions. These high-$J$ transitions are believed to trace the denser and often thermally excited gas in the more centrally concentrated in starburst/AGN regions. Our results that the CO(1-0) emission in HzRGs is often spread across tens of kpc scales thus strengthen the idea discussed in Sect. \ref{sec:intro} that studies of low- and high-$J$ transitions may be biased towards tracing different reservoirs of molecular gas. Thus, while a direct comparison between our study and the earlier work by \citet{oji97} and \citet{eva96} is difficult, it shows the need for sensitive studies of HzRGs across a wide range of molecular transitions and species.

\subsubsection{Low-$J$ transitions at high-$z$: a comparison}

\begin{figure}
\centering
\includegraphics[width=0.47\textwidth]{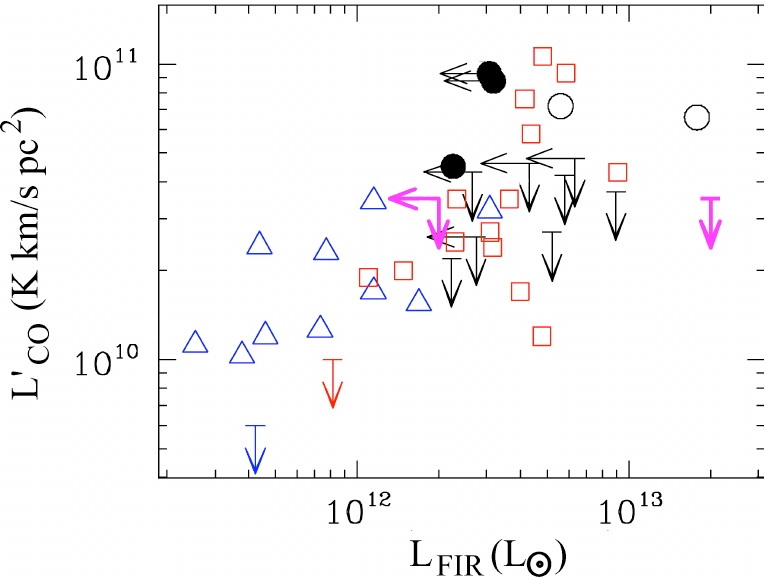}
\caption{$L'_{\rm CO}$ plotted against the starburst far-IR luminosity $L_{\rm FIR}$ for the various samples of high-$z$ galaxies observed in CO(1-0) or CO(2-1). HzRGs (black symbols) are as in Fig.\,\ref{fig:fir}. Starforming and `BzK'-selected galaxies (blue triangles) are taken from \citet{dad08} and \citet{ara11,ara12} (see also \citealt{dan09}). Submillimetre Galaxies (red squares + red arrow) are taken from \citet{ivi11} and \citet{bot12} (a SMG observed in CO(2-1) by \citealt{cop08} has been omitted due to a significant uncertainty in $L_{\rm FIR}$). CO(1-0) upper limits on two obscured QSOs (magenta arrows) are taken from \citet{lac11}. For targets observed in CO(2-1) the ratio $L'_{\rm CO(2-1)}$/$L'_{\rm CO(1-0)}$ adopted by the original studies was used to transform to CO(1-0) values, though corrections are small.}
\label{fig:samples}
\end{figure}

Our survey is the first survey for the ground-transition CO(1-0) in a representative sample of HzRGs. With upgrades to wide-band receivers and observing capabilities below $50$\,GHz at the large millimetre facilities, a growing number of studies are targeting the low-order CO transitions in various samples of high-$z$ galaxies. This allows a fair comparison of the cold molecular gas content of these various types of galaxies, without having to deal with uncertainty in the thermalisation of the gas or the possibility of missing CO emission due to the very limited velocity coverage that plagued past studies (see Sect.\,\ref{sec:intro}).

In Fig.\,\ref{fig:samples} we compare the CO(1-0) content of our HzRGs with that of samples of other types of galaxies in the same redshift range as our sample sources ($z \sim 1.5 - 2.5$). We only included studies performed with wideband receivers (covering $\geq 2000$\,\kms) that targeted either the CO(1-0) or CO(2-1) transition, resulting in comparison samples of submillimetre galaxies \citep[SMGs;][]{ivi11,hai11,bot12}, starforming galaxies \citep{dad08,dan09,ara10,ara12} and two obscured (type-2) quasi-stellar objects \citep[QSO2;][]{lac11}. Low-order CO observations done with narrow-band receivers \citep[e.g.][]{pap01,gre04,gre05} or observations that targeted either systems at $z>2.5$ \citep[e.g.][]{kla05,cop10,car11,wan11,ivi08,ivi12,huy13,rie13} or lensed systems \citep[e.g.][]{swi10,dan11,rie11,joh12,ara13lens,raw13} are excluded from our high-$z$ comparison in order to introduce as little bias as possible.

Figure\,\ref{fig:samples} shows $L'_{\rm CO(1-0)}$ plotted against the starburst far-IR luminosity $L_{\rm FIR}$ for the various types of high-$z$ galaxies.\footnote{It is likely that $L_{\rm FIR}$ in some of the SMGs and QSO2 that we present is contaminated by an AGN-contribution (unlike for the HzRG, where the AGN contribution is separated from the starburst $L_{\rm FIR}$ values; Sect.\,\ref{sec:cofir}). However we argue that this will not likely alter the main conclusions that we derive in this Section.} It is immediately clear that the five CO-detected HzRGs in our sample have a CO(1-0) luminosity that is comparable to what is found in the CO-brightest SMGs. 

Our CO(1-0) detection rate of 38$\%$ (taking into consideration a typical 3$\sigma$ detection limit of $L`_{\rm CO(1-0)} \la 4 \times 10^{10}$\,K\,\kms\,pc$^{2}$) is in rough agreement with the fact that 33$\%$ of the SMGs observed in CO(1-0) or CO(2-1) have $L`_{\rm CO(1-0)} \geq 4 \times 10^{10}$\,K\,\kms\,pc$^{2}$ (Fig.\,\ref{fig:samples}). None of the $z \sim 1.5 - 2.5$ starforming galaxies is detected in CO(1-0) at this level, which reflects in their lower $L_{\rm FIR}$.  

It is, however, interesting that when excluding the three off-nuclear CO(1-0) detections in our sample (i.e. when only taking into account CO detections at the location of the host galaxy), the CO(1-0) detection rate of HzRGs drops to less than half that of SMGs (assuming a similar sensitivity cutoff). It would be worth to verify a possible deficiency of CO(1-0) emission in HzRGs, and investigate whether this can be related to the observed shock-heating of molecular H$_{2}$ gas in the centres of HzRGs \citep{nes10,ogl12,gui12} and the highly shock-excited (`above thermalisation') CO-emitting gas found in some radio galaxies \citep[][]{pap08293,ivi12}.

Larger CO surveys of unbiased samples with EVLA and ALMA are required to further investigate how the amount and properties of the CO-emitting gas in HzRGs compares to that of other types of high-$z$ galaxies.

\section{Conclusions}
\label{sec:conclusions}

We have performed the first representative survey for cold molecular CO(1-0) gas in a sample of 13 high-z radio galaxies ($1.4<z<2.8$) using the Australia Telescope Compact Array. The main results from this work are:
\vspace{2mm}\\
{\sl i).} We detect CO(1-0) emission in 38$\%$ (5/13) of our sample sources. The CO(1-0) luminosities are in the range $L'_{\rm CO} = (4.5 - 9.2) \times 10^{10}$\,K\,\kms\,pc$^{2}$, which correspond to molecular gas masses of M$_{\rm H2} = (4 - 7) \times 10^{10}\, {\rm M}_{\odot}$ when assuming M$_{\rm H2}$/$L'_{\rm CO}$\,=\,0.8. The CO(1-0) profile for four of the five detections is broad (FWZI $\sim$ 1100 - 3600 \kms).\\
\vspace{-1mm}\\
{\sl ii).} Only for two of these sources (MRC\,0152-209 and MRC\,1138-262), part of the CO(1-0) emission is co-spatial with the radio host galaxy, although a significant fraction of the CO(1-0) is spread across tens of kpc scales and likely related to the process of galaxy merging.\\
\vspace{-1mm}\\
{\sl iii).} For the other three CO-detections (MRC\,0114-211, MRC\,0156-252 and MRC\,2048-272) the CO(1-0) is found in the halo of the host galaxy. These three HzRG are among the fainter FIR emitters in the sample, indicating that large amounts of molecular halo gas ($L'_{\rm CO(1-0)}$= several $\times 10^{10}$\,K\,\kms\,pc$^{2}$) may have been missed in previous studies that mostly pre-selected targets based on a high FIR luminosity.\\
\vspace{-1mm}\\
{\sl iv).} We find an alignment between the off-nuclear CO(1-0) emission and the radio jet axis, with the CO located outside the brightest edge of the radio source. We discuss several scenarios that may explain this, including jet-induced star formation / gas cooling, jet-driven metal-enrichment of the gaseous halo and flux boosting of jets that propagating into a dense filament of cold halo gas.\\
\vspace{-1mm}\\
{\sl v).} Following our results, we performed a literature study on extended CO($J,J-1$) emission in HzRG and find marginal statistical significance (95$\%$ level) for the hypothesis that the CO is preferentially aligned towards to the radio axis.\\
\vspace{-1mm}\\
{\sl vi).} The majority of the host galaxies of high-$z$ radio sources does not contain CO(1-0) emission down to a secure limit of $L'_{\rm CO} \sim 3-4 \times 10^{10}$\,K\,\kms\,pc$^{2}$.
\vspace{2mm}\\
\indent We have shown that from now on, the well-known alignments between the radio jets and optical, UV, X-ray and submm emission in HzRGs (see Sect. \ref{sec:intro}) will have to be discussed taking into account the component of cold molecular gas. 

We note that the CO(1-0) emission that we found in the halos of three HzRG may only reveal part of the cold molecular gas content around HzRGs, as our observations are are only sensitive to tracing those sites where the CO emission is brightest. Follow-up observations with the ATCA, EVLA and ALMA are essential to confirm the general existence of large reservoirs of cold molecular halo gas and study them in more detail. Given that the CO in these halo reservoirs may be widespread and is likely not forming stars at the rates seen in high-$z$ starforming and submillimetre galaxies, we argue that it is important to target the low-$J$ transitions of CO with short-baseline array configurations, and to study samples that are unbiased in IR or submm luminosity.

\section*{Acknowledgments}
A warm thanks to the dedicated staff at the Australia Telescope Compact Array in Narrabri for their round-the-clock support, with a special thanks to Robin Wark, Warwick Wilson, Jamie Stevens, Phil Edwards and Mark Wieringa for their extra efforts regarding the observing, scheduling and software updates that made this work possible. We also thank Santiago Arribas for useful feedback while drafting this paper and Nina Hatch for her help in trying to retrieve archived data. Thanks to the anonymous referee for suggestions that significantly improved this paper. BE thanks the Centro de Astrobiolog\'{i}a (CSIC/INTA) for their hospitality, and CSIRO Astronomy and Space Science for their generousity, during the final stage of this project. The Australia Telescope is funded by the Commonwealth of Australia for operation as a National Facility managed by CSIRO. This research has made use of the NASA/IPAC Extragalactic Database (NED) which is operated by the Jet Propulsion Laboratory, California Institute of Technology, under contract with the National Aeronautics and Space Administration. Some of the results presented in this paper were based on observations made with the NASA/ESA Hubble Space Telescope, and obtained from the Hubble Legacy Archive, which is a collaboration between the Space Telescope Science Institute (STScI/NASA), the Space Telescope European Coordinating Facility (ST-ECF/ESA) and the Canadian Astronomy Data Centre (CADC/NRC/CSA). BE acknowledges funding from the Australian Commonwealth Scientific and Industrial Research Organisation (CSIRO) and additional support by the Spanish Ministerio de Econom\'{i}a y Competitividad under grant AYA2010-21161-C02-01. NS is a recipient of an ARC Future Fellowship.

\appendix

\section{Notes on individual objects and their CO(1-0) content}
\label{sec:app}

\noindent{\bf MRC\,0114-211 ($z=1.40$):} MRC\,0114-211 has a strong  ($\sim80$\,mJy) 48\,GHz radio continuum flux, and was also detected in the Australia Telescope 20\,GHz (AT20G) Survey \citep{mur10,mas11}. The radio source is a Compact Steep Spectrum object with a total linear extent of 6 kpc \citep{bre10,ran11}. 

We detect CO(1-0) emission 4 arcsec (34 kpc) west of the radio core, along the axis of the compact radio jet (Fig. \ref{fig:COdistribution}). An archival {\sl HST} image from Program 8838 (PI: Lehnert) obtained with the Wide-Field Planetary Camera 2 (WFPC2$_{\rm F555W}$) shows faint optical emission centred at the location of the CO(1-0) emission, with brighter emission found just outside the edge of the radio source (Fig.\,\ref{fig:app0114}). However, since the archival {\sl HST} data were calibrated using the Guide Star Catalogue 1 \citep[GSC-1;][]{las90}, an astrometric error of $\sim1.5$'' in these {\sl HST} data is possible \citep{rus88,rus90}, which could place the bright optical emission at the location of the host galaxy.

As shown in Table \ref{tab:obs}, the observing time on MRC\,0114-211 had to be longer than for the other sources in our sample to reach a similar sensitivity (because observations were done in the lower sensitivity part of the ATCA 7mm band). In total 7 independent runs of 3\,-\,5 hours on-source observing time each were devoted to this target. Low-level indications for the CO(1-0) signal are present across the individual data sets from these runs. 

MRC\,0114-211 is the only source in our sample with a large uncertainty in optical redshift ($z = 1.41 \pm 0.05$; \citealt{mcc96})\\
\ \\
\noindent{\bf MRC\,0152-209 ($z=1.92$):} MRC\,0152-209 has an HST morphology reminiscent of a major merger system, with large-scale tidal tails and a double nucleus \citep[][Emonts et al in prep.]{pen01}. The 18\,kpc-wide radio source is aligned along the main optical emission from the host galaxy \citep{pen00radio,pen01}. 

The CO(1-0) detection is discussed in detail in \citet{emo11a}. Due to the relatively narrow FWHM $\approx$ 400 \kms\ of the CO profile, the CO(1-0) peaks at high significance and a high-resolution follow-up study to spatially map the CO emission is in progress (Emonts et al in prep; see also Fig.\,\ref{fig:mrc0152color}).\\
\ \\
\noindent {\bf MRC\,0156-252 ($z=2.02$):} MRC\,0156-252 contains an extended radio source \citep[diameter 70 kpc;][]{car97}. \citet{pen01} showed that there are several companion galaxies detected with {\sl HST/NICMOS} that are aligned along the radio axis, as well as a large reservoir of \lya\ emission that stretches inside the extent of the radio source.

The CO(1-0) emission in MRC\,0156-209 peaks just outside the bright hot-spot of the bent NE radio jet ($\sim$50 kpc or 6 arcsec from the radio core) and aligns with the reservoir of \lya\ emission. It is thus feasible that the CO(1-0) and \lya\ emission trace the same gas reservoir, which is ionised within the extent of the radio source \citep[possibly as a combination of photo-ionisation of the AGN and shock-excitation by the radio jet;][]{vil03,ove05}. Just north of the hot-spot and region of \lya\ and CO(1-0) emission, \citet{ove05} found a tentative reservoir of diffuse X-ray emission. Therefore, an even more complex gas reservoir (with a wide range of temperatures) may surround the bright radio lobe structure.

Figure \ref{fig:mrc0156lyaco} shows that the CO(1-0) emission appears to consist of two marginally resolved components that are located less than 15-20 kpc from each of the two NE companions in the {\sl HST/NICMOS} image. Recent work by \citet{gal13} suggests that the redshift at which we centred our CO(1-0) profile ($z=2.016$) is closely related to the redshift of the innermost companion ($z_{\rm He\,II} = 2.0171 \pm 0.0004$), which has blue colors indicative of star formation \citep[][see also Sect.\,\ref{sec:off}]{pen01}. According to \citet{gal13}, this blue companion is shifted by almost $-1000$ \kms\ with respect to the revised redshift of the central HzRG ($z_{\rm He\,II} = 2.0256 \pm 0.0002$). The `blue' and `red' CO(1-0) component in Fig.\,\ref{fig:mrc0156lyaco} have a luminosity of $L'_{\rm CO} \sim 5 \times 10^{10}$  and $\sim 4 \times 10^{10}$ K\,\kms\,pc$^{2}$, respectively.

\begin{figure}
\centering
\includegraphics[width=0.45\textwidth]{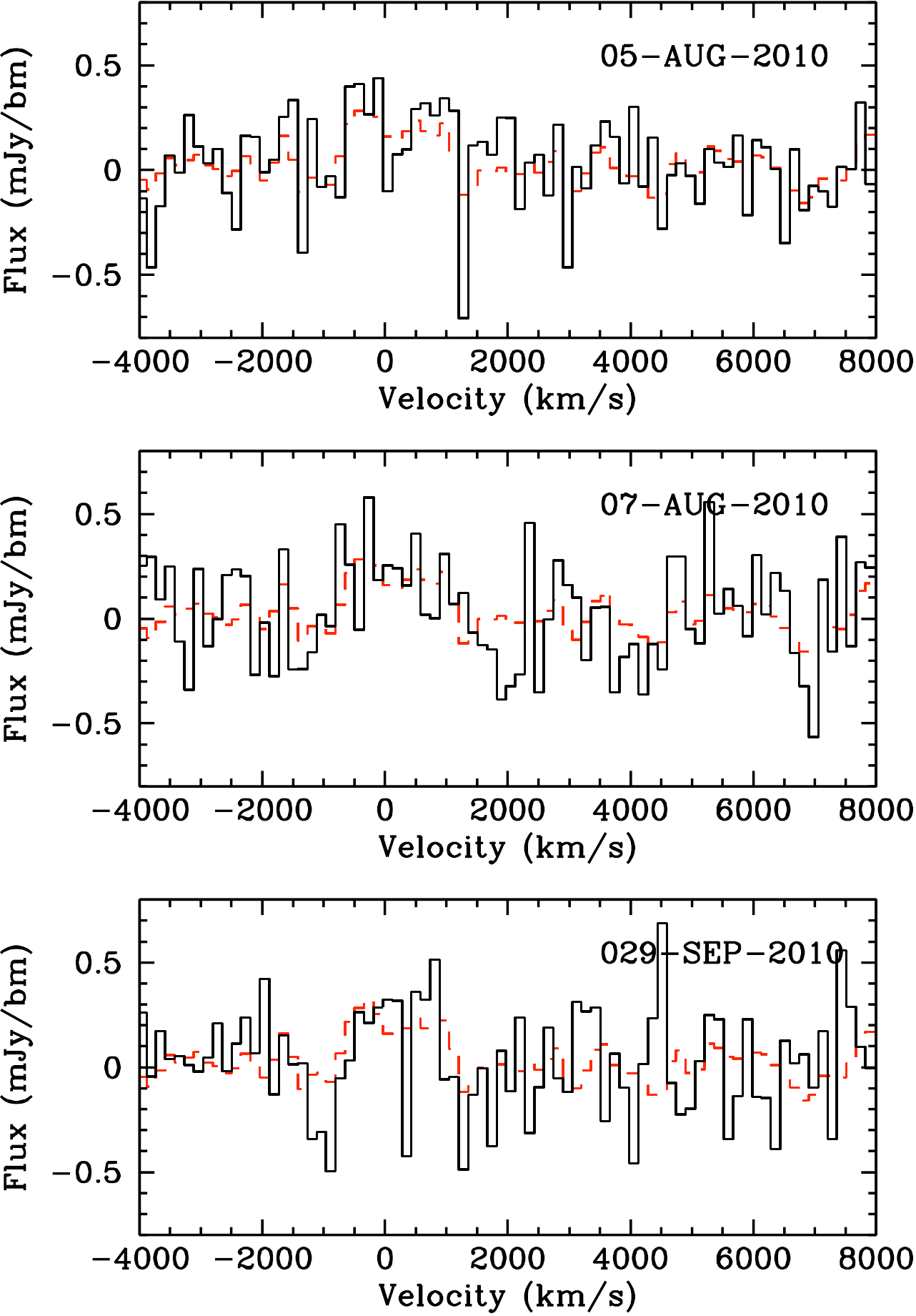}
\caption{CO(1-0) profiles of MRC\,0156-252 taken at the three difference observing epochs (Table \ref{tab:obs}). The dashed red line is the total spectrum from Fig. \ref{fig:COprofiles}.}
\label{fig:mrc0156}
\end{figure}

Because MRC\,0156-252 was targeted only during three observing runs, Fig. \ref{fig:mrc0156} shows the spectra of the three individual observing epochs. Indications for the signal are present at low level significance in the data of all three epochs. New observations are in progress to better map the CO(1-0) emission.\\
\ \\
\noindent{\bf MRC\,1138-262 ($z=2.16$):} MRC\,1138-262, also called the Spiderweb Galaxy, is one of the most massive and active galaxies in the Early Universe \citep{mil06,sey07,bre10}. It is a conglomerate of star-forming galaxies that are embedded in a giant ($>200$\,kpc) Ly$\alpha$ halo, located in the core of the Spiderweb proto-cluster \citep{pen97,pen00,pen02,car98,kur04,cro05,kod07,zir08,hat09,doh10,kui11}. Massive star formation (SFR$\sim$1400\,M$_{\odot}$\,yr$^{-1}$) occurs on scales of $>200$\,kpc \citep{ste03,sey07,ogl12}. The central galaxy hosts the radio source MRC\,1138-262, which induces significant feedback onto the surrounding ISM \citep{nes06,ogl12}.

Our CO(1-0) results are discussed in \citet{emo13}. Part of the CO(1-0) emission is associated with the central host galaxy of the radio source, but a significant fraction of the gas is spread across at least several tens of kpc (Fig.\,\ref{fig:mrc1138color}), most likely associated with merging companion galaxies or the IGM between them \citep[see][for a discussion]{emo13}. \\
\ \\
\noindent{\bf MRC\,2048-272 ($z=2.06$):} {\sl HST/NICMOS} and VLA imaging by \citet{pen00,pen01} revealed that the radio source is likely hosted by a bright NIR object that is surrounded by two NIR companions. The brighter of the two companions is an AGN \citep{bre10}. The fainter of the two NIR companions is located in between the tip of the bright NE radio hot-spot and the location where the CO(1-0) peaks. Another IR-bright source is located $\sim8$ arcsec SE of the radio galaxy \citep{bre10}. 

For MRC\,2048-272 the CO(1-0) profile appears double-peaked and centred around $z_{\rm Ly\alpha}$ from \citet{ven07}. While the overall CO(1-0) emission covers a very wide velocity range (FWZI = 3570 \kms), both peaks are spatially unresolved and centred at the same location, about 56 kpc or 6 arcsec (i.e. one synthesized beam radius) away from the core of the radio galaxy. For both the blue and red peak $L'_{\rm CO} \sim 4 \times 10^{10}$\,K\,\kms\,pc$^{2}$. More sensitive observations are required to verify the exact shape of the CO profile and related total CO(1-0) intensity.


\end{document}